\documentclass{article}
\usepackage[a4paper,margin=1.3in]{geometry}
\pdfoutput=1
\usepackage[utf8]{inputenc}
\usepackage[english]{babel}
\usepackage[T1]{fontenc}
\usepackage{amsmath,amssymb,amsfonts,thmtools,thm-restate,mathtools,physics, mathrsfs, enumitem, braket}
\usepackage[round]{natbib}
\usepackage{csquotes}

\usepackage[dvipsnames]{xcolor}
\usepackage[pdfusetitle,bookmarksnumbered,colorlinks,linkcolor=MidnightBlue,citecolor=olive]{hyperref}
\usepackage[capitalise,nameinlink,noabbrev]{cleveref}
\crefname{equation}{Eq.}{Eqs.}
\Crefname{equation}{Eq.}{Eqs.}
\crefname{assumption}{Assumption}{Assumptions}
\Crefname{assumption}{Assumption}{Assumptions}
\usepackage{tikzit}
\usepackage{tikz, circuitikz}
\tikzstyle{env}=[copoint,regular polygon rotate=0,minimum width=0.2cm, fill=black]

\tikzstyle{probs}=[shape=semicircle,fill=white,draw=black,shape border rotate=180,minimum width=1.2cm]

\tikzstyle{nudge}=[yshift=0.6mm]

\tikzstyle{every picture}=[baseline=-0.25em,scale=0.5]
\tikzstyle{dotpic}=[] 
\tikzstyle{diredges}=[every to/.style={diredge}]
\tikzstyle{math matrix}=[matrix of math nodes,left delimiter=(,right delimiter=),inner sep=2pt,column sep=1em,row sep=0.5em,nodes={inner sep=0pt},text height=1.5ex, text depth=0.25ex]

\tikzstyle{inline text}=[text height=1.5ex, text depth=0.25ex,yshift=0.5mm]
\tikzstyle{label}=[font=\footnotesize,text height=1.5ex, text depth=0.25ex]
\tikzstyle{left label}=[label,anchor=east,xshift=2mm]
\tikzstyle{right label}=[label,anchor=west,xshift=-2mm]

\tikzstyle{braceedge}=[decorate,decoration={brace,amplitude=2mm,raise=-1mm}]
\tikzstyle{small braceedge}=[decorate,decoration={brace,amplitude=1mm,raise=-1mm}]

\tikzstyle{doubled}=[line width=1.6pt] 
\tikzstyle{boldedge}=[doubled,shorten <=-0.17mm,shorten >=-0.17mm]
\tikzstyle{boldedgegray}=[doubled,gray,shorten <=-0.17mm,shorten >=-0.17mm]
\tikzstyle{singleedgegray}=[gray]

\tikzstyle{semidoubled}=[line width=1.4pt] 
\tikzstyle{semiboldedgegray}=[semidoubled,gray,shorten <=-0.17mm,shorten >=-0.17mm]

\tikzstyle{boxedge}=[semiboldedgegray]

\tikzstyle{boldedgedashed}=[very thick,dashed,shorten <=-0.17mm,shorten >=-0.17mm]
\tikzstyle{vboldedgedashed}=[doubled,dashed,shorten <=-0.17mm,shorten >=-0.17mm]
\tikzstyle{left hook arrow}=[left hook-latex]
\tikzstyle{right hook arrow}=[right hook-latex]
\tikzstyle{sembracket}=[line width=0.5pt,shorten <=-0.07mm,shorten >=-0.07mm]

\tikzstyle{causal edge}=[->,thick,gray]
\tikzstyle{causal nondir}=[thick,gray]
\tikzstyle{timeline}=[thick,gray, dashed]

\tikzstyle{cedge}=[<->,thick,gray!70!white]

\tikzstyle{empty diagram}=[draw=gray!40!white,dashed,shape=rectangle,minimum width=1cm,minimum height=1cm]
\tikzstyle{empty diagram small}=[draw=gray!50!white,dashed,shape=rectangle,minimum width=0.6cm,minimum height=0.5cm]

\tikzstyle{dot}=[inner sep=0mm,minimum width=2mm,minimum height=2mm,draw,shape=circle]  
\tikzstyle{Wsquare}=[white dot, shape=regular polygon, rounded corners=0.8 mm, minimum size=3.3 mm, regular polygon sides=3, outer sep=-0.2mm]
\tikzstyle{Wsquareadj}=[white dot, shape=regular polygon, rounded corners=0.8 mm, minimum size=3.3 mm, regular polygon sides=3, outer sep=-0.2mm, regular polygon rotate=180]
\tikzstyle{ddot}=[inner sep=0mm, doubled, minimum width=2.5mm,minimum height=2.5mm,draw,shape=circle]

\tikzstyle{black dot}=[dot,fill=black]
\tikzstyle{white dot}=[dot,fill=white,,text depth=-0.2mm]
\tikzstyle{white Wsquare}=[Wsquare,fill=white,,text depth=-0.2mm]
\tikzstyle{white Wsquareadj}=[Wsquareadj,fill=white,,text depth=-0.2mm]
\tikzstyle{green dot}=[white dot] 
\tikzstyle{gray dot}=[dot,fill=gray!40!white,,text depth=-0.2mm]
\tikzstyle{red dot}=[gray dot] 

\tikzstyle{black ddot}=[ddot,fill=black]
\tikzstyle{white ddot}=[ddot,fill=white]
\tikzstyle{gray ddot}=[ddot,fill=gray!40!white]

\tikzstyle{gray edge}=[gray!60!white]

\tikzstyle{small dot}=[inner sep=0.5mm,minimum width=0pt,minimum height=0pt,draw,shape=circle]

\tikzstyle{small black dot}=[small dot,fill=black]
\tikzstyle{small white dot}=[small dot,fill=white]
\tikzstyle{small gray dot}=[small dot,fill=gray!40!white]

\tikzstyle{very small dot}=[inner sep=0.3mm,minimum width=0pt,minimum height=0pt,draw,shape=circle]

\tikzstyle{very small black dot}=[very small dot,fill=black]
\tikzstyle{very small white dot}=[small dot,fill=white]
\tikzstyle{very small gray dot}=[small dot,fill=gray!40!white]

\tikzstyle{causal dot}=[inner sep=0.4mm,minimum width=0pt,minimum height=0pt,draw=white,shape=circle,fill=gray!40!white]

\tikzstyle{phase dimensions}=[minimum size=5mm,font=\footnotesize,rectangle,rounded corners=2.5mm,inner sep=0.2mm,outer sep=-2mm]
\tikzstyle{dphase dimensions}=[minimum size=5mm,font=\footnotesize,rectangle,rounded corners=2.5mm,inner sep=0.2mm,outer sep=-2mm]

\tikzstyle{white phase dot}=[dot,fill=white,phase dimensions]
\tikzstyle{white phase ddot}=[ddot,fill=white,dphase dimensions]

\tikzstyle{white rect ddot}=[draw=black,fill=white,doubled,minimum size=5mm,font=\footnotesize,rectangle,rounded corners=2.5mm,inner sep=0.2mm]
\tikzstyle{gray rect ddot}=[draw=black,fill=gray!40!white,doubled,minimum size=6mm,font=\footnotesize,rectangle,rounded corners=3mm]

\tikzstyle{gray phase dot}=[dot,fill=gray!40!white,phase dimensions]
\tikzstyle{gray phase ddot}=[ddot,fill=gray!40!white,dphase dimensions]
\tikzstyle{grey phase dot}=[gray phase dot]
\tikzstyle{grey phase ddot}=[gray phase ddot]

\tikzstyle{small phase dimensions}=[minimum size=4mm,font=\tiny,rectangle,rounded corners=2mm,inner sep=0.2mm,outer sep=-2mm]
\tikzstyle{small dphase dimensions}=[minimum size=4mm,font=\tiny,rectangle,rounded corners=2mm,inner sep=0.2mm,outer sep=-2mm]

\tikzstyle{small gray phase dot}=[dot,fill=gray!40!white,small phase dimensions]
\tikzstyle{small gray phase ddot}=[ddot,fill=gray!40!white,small dphase dimensions]

\tikzstyle{small map}=[draw,shape=rectangle,minimum height=4mm,minimum width=4mm,fill=white]

\tikzstyle{cnot}=[fill=white,shape=circle,inner sep=-1.4pt]

\tikzstyle{asym hadamard}=[fill=white,draw,shape=NEbox,inner sep=0.6mm,font=\footnotesize,minimum height=4mm]
\tikzstyle{asym hadamard conj}=[fill=white,draw,shape=NWbox,inner sep=0.6mm,font=\footnotesize,minimum height=4mm]
\tikzstyle{asym hadamard dag}=[fill=white,draw,shape=SEbox,inner sep=0.6mm,font=\footnotesize,minimum height=4mm]

\tikzstyle{hadamard}=[fill=white,draw,inner sep=0.6mm,font=\footnotesize,minimum height=4mm,minimum width=4mm]
\tikzstyle{small hadamard}=[fill=white,draw,inner sep=0.6mm,minimum height=1.5mm,minimum width=1.5mm]
\tikzstyle{small hadamard rotate}=[small hadamard,rotate=45]
\tikzstyle{dhadamard}=[hadamard,doubled]
\tikzstyle{small dhadamard}=[small hadamard,doubled]
\tikzstyle{small dhadamard rotate}=[small hadamard rotate,doubled]
\tikzstyle{antipode}=[white dot,inner sep=0.3mm,font=\footnotesize]

\tikzstyle{scalar}=[diamond,draw,inner sep=0.5pt,font=\small]
\tikzstyle{dscalar}=[diamond,doubled, draw,inner sep=0.5pt,font=\small]

\tikzstyle{small box}=[rectangle,inline text,fill=white,draw,minimum height=5mm,yshift=-0.5mm,minimum width=5mm,font=\small]
\tikzstyle{small gray box}=[small box,fill=gray!30]
\tikzstyle{medium box}=[rectangle,inline text,fill=white,draw,minimum height=5mm,yshift=-0.5mm,minimum width=8mm,font=\small]
\tikzstyle{square box}=[small box] 
\tikzstyle{medium gray box}=[small box,fill=gray!30]
\tikzstyle{semilarge box}=[rectangle,inline text,fill=white,draw,minimum height=5mm,yshift=-0.5mm,minimum width=12.5mm,font=\small]
\tikzstyle{large box}=[rectangle,inline text,fill=white,draw,minimum height=5mm,yshift=-0.5mm,minimum width=15mm,font=\small]
\tikzstyle{large gray box}=[small box,fill=gray!30]

\tikzstyle{Bayes box}=[rectangle,fill=black,draw, minimum height=3mm, minimum width=3mm]

\tikzstyle{gray square point}=[small box,fill=gray!50]

\tikzstyle{dphase box white}=[dhadamard]
\tikzstyle{dphase box gray}=[dhadamard,fill=gray!50!white]
\tikzstyle{phase box white}=[hadamard]
\tikzstyle{phase box gray}=[hadamard,fill=gray!50!white]

\tikzstyle{point}=[regular polygon,regular polygon sides=3,draw,scale=0.75,inner sep=-0.5pt,minimum width=9mm,fill=white,regular polygon rotate=180]
\tikzstyle{copoint}=[regular polygon,regular polygon sides=3,draw,scale=0.75,inner sep=-0.5pt,minimum width=9mm,fill=white]
\tikzstyle{dpoint}=[point,doubled]
\tikzstyle{dcopoint}=[copoint,doubled]

\tikzstyle{wide copoint}=[fill=white,draw,shape=isosceles triangle,shape border rotate=90,isosceles triangle stretches=true,inner sep=0pt,minimum width=1.5cm,minimum height=6.12mm]
\tikzstyle{wide point}=[fill=white,draw,shape=isosceles triangle,shape border rotate=-90,isosceles triangle stretches=true,inner sep=0pt,minimum width=1.5cm,minimum height=6.12mm,yshift=-0.0mm]
\tikzstyle{wide point plus}=[fill=white,draw,shape=isosceles triangle,shape border rotate=-90,isosceles triangle stretches=true,inner sep=0pt,minimum width=1.74cm,minimum height=7mm,yshift=-0.0mm]

\tikzstyle{wide dpoint}=[fill=white,doubled,draw,shape=isosceles triangle,shape border rotate=-90,isosceles triangle stretches=true,inner sep=0pt,minimum width=1.5cm,minimum height=6.12mm,yshift=-0.0mm]

\tikzstyle{tinypoint}=[regular polygon,regular polygon sides=3,draw,scale=0.55,inner sep=-0.15pt,minimum width=6mm,fill=white,regular polygon rotate=180] 

\tikzstyle{white point}=[point]
\tikzstyle{white dpoint}=[dpoint]
\tikzstyle{green point}=[white point] 
\tikzstyle{white copoint}=[copoint]
\tikzstyle{gray point}=[point,fill=gray!40!white]
\tikzstyle{gray dpoint}=[gray point,doubled]
\tikzstyle{red point}=[gray point] 
\tikzstyle{gray copoint}=[copoint,fill=gray!40!white]
\tikzstyle{gray dcopoint}=[gray copoint,doubled]

\tikzstyle{white point guide}=[regular polygon,regular polygon sides=3,font=\scriptsize,draw,scale=0.65,inner sep=-0.5pt,minimum width=9mm,fill=white,regular polygon rotate=180]

\tikzstyle{black point}=[point,fill=black,font=\color{white}]
\tikzstyle{black copoint}=[copoint,fill=black,font=\color{white}]

\tikzstyle{tiny gray point}=[tinypoint,fill=gray!40!white]

\tikzstyle{diredge}=[->]
\tikzstyle{ddiredge}=[<->]
\tikzstyle{rdiredge}=[<-]
\tikzstyle{thickdiredge}=[->, very thick]
\tikzstyle{pointer edge}=[->,very thick,gray]
\tikzstyle{pointer edge part}=[very thick,gray]
\tikzstyle{dashed edge}=[dashed]
\tikzstyle{thick dashed edge}=[very thick,dashed]
\tikzstyle{thick gray dashed edge}=[thick dashed edge,gray!40]
\tikzstyle{thick map edge}=[very thick,|->]

\makeatletter
\newcommand{\boxshape}[3]{%
\pgfdeclareshape{#1}{
\inheritsavedanchors[from=rectangle] 
\inheritanchorborder[from=rectangle]
\inheritanchor[from=rectangle]{center}
\inheritanchor[from=rectangle]{north}
\inheritanchor[from=rectangle]{south}
\inheritanchor[from=rectangle]{west}
\inheritanchor[from=rectangle]{east}
\backgroundpath{
\southwest \pgf@xa=\pgf@x \pgf@ya=\pgf@y
\northeast \pgf@xb=\pgf@x \pgf@yb=\pgf@y

\@tempdima=#2
\@tempdimb=#3

\pgfpathmoveto{\pgfpoint{\pgf@xa - 5pt + \@tempdima}{\pgf@ya}}
\pgfpathlineto{\pgfpoint{\pgf@xa - 5pt - \@tempdima}{\pgf@yb}}
\pgfpathlineto{\pgfpoint{\pgf@xb + 5pt + \@tempdimb}{\pgf@yb}}
\pgfpathlineto{\pgfpoint{\pgf@xb + 5pt - \@tempdimb}{\pgf@ya}}
\pgfpathlineto{\pgfpoint{\pgf@xa - 5pt + \@tempdima}{\pgf@ya}}
\pgfpathclose
}
}}

\boxshape{NEbox}{0pt}{0pt}
\boxshape{SEbox}{0pt}{-5pt}
\boxshape{NWbox}{5pt}{0pt}
\boxshape{SWbox}{-5pt}{0pt}
\boxshape{EBox}{-3pt}{3pt}
\boxshape{WBox}{3pt}{-3pt}
\makeatother

\tikzstyle{cloud}=[shape=cloud,draw,minimum width=1.5cm,minimum height=1.5cm]

\tikzstyle{map}=[draw,shape=NEbox,inner sep=2pt,minimum height=6mm,fill=white]
\tikzstyle{dashedmap}=[draw,dashed,gray,shape=NEbox,inner sep=2pt,minimum height=6mm,fill=white]
\tikzstyle{medium dashedmap}=[draw,dashed,gray,shape=NEbox,inner sep=2pt,minimum height=6mm,fill=white,minimum width=7mm]
\tikzstyle{semilarge dashedmap}=[draw,dashed,gray,shape=NEbox,inner sep=2pt,minimum height=6mm,fill=white,minimum width=9.5mm]
\tikzstyle{large dashedmap}=[draw,dashed,gray,shape=NEbox,inner sep=2pt,minimum height=6mm,fill=white,minimum width=12mm]
\tikzstyle{very large dashedmap}=[draw,dashed,gray,shape=NEbox,inner sep=2pt,minimum height=6mm,fill=white,minimum width=17mm]

\tikzstyle{dashed map}=[fill=white, draw=gray, shape=rectangle, style=map, dashed]

\tikzstyle{mapdag}=[draw,shape=SEbox,inner sep=2pt,minimum height=6mm,fill=white]
\tikzstyle{mapadj}=[draw,shape=SEbox,inner sep=2pt,minimum height=6mm,fill=white]
\tikzstyle{maptrans}=[draw,shape=SWbox,inner sep=2pt,minimum height=6mm,fill=white]
\tikzstyle{mapconj}=[draw,shape=NWbox,inner sep=2pt,minimum height=6mm,fill=white]

\tikzstyle{medium map}=[draw,shape=NEbox,inner sep=2pt,minimum height=6mm,fill=white,minimum width=7mm]
\tikzstyle{medium map dag}=[draw,shape=SEbox,inner sep=2pt,minimum height=6mm,fill=white,minimum width=7mm]
\tikzstyle{medium map adj}=[draw,shape=SEbox,inner sep=2pt,minimum height=6mm,fill=white,minimum width=7mm]
\tikzstyle{medium map trans}=[draw,shape=SWbox,inner sep=2pt,minimum height=6mm,fill=white,minimum width=7mm]
\tikzstyle{medium map conj}=[draw,shape=NWbox,inner sep=2pt,minimum height=6mm,fill=white,minimum width=7mm]
\tikzstyle{semilarge map}=[draw,shape=NEbox,inner sep=2pt,minimum height=6mm,fill=white,minimum width=9.5mm]
\tikzstyle{semilarge map trans}=[draw,shape=SWbox,inner sep=2pt,minimum height=6mm,fill=white,minimum width=9.5mm]
\tikzstyle{semilarge map adj}=[draw,shape=SEbox,inner sep=2pt,minimum height=6mm,fill=white,minimum width=9.5mm]
\tikzstyle{semilarge map dag}=[draw,shape=SEbox,inner sep=2pt,minimum height=6mm,fill=white,minimum width=9.5mm]
\tikzstyle{semilarge map conj}=[draw,shape=NWbox,inner sep=2pt,minimum height=6mm,fill=white,minimum width=9.5mm]
\tikzstyle{large map}=[draw,shape=NEbox,inner sep=2pt,minimum height=6mm,fill=white,minimum width=12mm]
\tikzstyle{large map conj}=[draw,shape=NWbox,inner sep=2pt,minimum height=6mm,fill=white,minimum width=12mm]
\tikzstyle{very large map}=[draw,shape=NEbox,inner sep=2pt,minimum height=6mm,fill=white,minimum width=17mm]
\tikzstyle{very very large map}=[draw,shape=NEbox,inner sep=2pt,minimum height=6mm,fill=white,minimum width=50mm]
\tikzstyle{large map dag}=[draw,shape=SEbox,inner sep=2pt,minimum height=6mm,fill=white,minimum width=12mm]

\tikzstyle{medium dmap}=[draw,doubled,shape=NEbox,inner sep=2pt,minimum height=6mm,fill=white,minimum width=7mm]
\tikzstyle{medium dmap dag}=[draw,doubled,shape=SEbox,inner sep=2pt,minimum height=6mm,fill=white,minimum width=7mm]
\tikzstyle{medium dmap adj}=[draw,doubled,shape=SEbox,inner sep=2pt,minimum height=6mm,fill=white,minimum width=7mm]
\tikzstyle{medium dmap trans}=[draw,doubled,shape=SWbox,inner sep=2pt,minimum height=6mm,fill=white,minimum width=7mm]
\tikzstyle{medium dmap conj}=[draw,doubled,shape=NWbox,inner sep=2pt,minimum height=6mm,fill=white,minimum width=7mm]
\tikzstyle{semilarge dmap}=[draw,doubled,shape=NEbox,inner sep=2pt,minimum height=6mm,fill=white,minimum width=9.5mm]
\tikzstyle{semilarge dmap trans}=[draw,doubled,shape=SWbox,inner sep=2pt,minimum height=6mm,fill=white,minimum width=9.5mm]
\tikzstyle{semilarge dmap adj}=[draw,doubled,shape=SEbox,inner sep=2pt,minimum height=6mm,fill=white,minimum width=9.5mm]
\tikzstyle{semilarge dmap dag}=[draw,doubled,shape=SEbox,inner sep=2pt,minimum height=6mm,fill=white,minimum width=9.5mm]
\tikzstyle{semilarge dmap conj}=[draw,doubled,shape=NWbox,inner sep=2pt,minimum height=6mm,fill=white,minimum width=9.5mm]
\tikzstyle{large dmap}=[draw,doubled,shape=NEbox,inner sep=2pt,minimum height=6mm,fill=white,minimum width=12mm]
\tikzstyle{large dmap conj}=[draw,doubled,shape=NWbox,inner sep=2pt,minimum height=6mm,fill=white,minimum width=12mm]
\tikzstyle{large dmap trans}=[draw,doubled,shape=SWbox,inner sep=2pt,minimum height=6mm,fill=white,minimum width=12mm]
\tikzstyle{large dmap adj}=[draw,doubled,shape=SEbox,inner sep=2pt,minimum height=6mm,fill=white,minimum width=12mm]
\tikzstyle{large dmap dag}=[draw,doubled,shape=SEbox,inner sep=2pt,minimum height=6mm,fill=white,minimum width=12mm]
\tikzstyle{very large dmap}=[draw,doubled,shape=NEbox,inner sep=2pt,minimum height=6mm,fill=white,minimum width=19.5mm]

\tikzstyle{muxbox}=[draw,shape=rectangle,minimum height=3mm,minimum width=3mm,fill=white]
\tikzstyle{dmuxbox}=[muxbox,doubled]

\tikzstyle{box}=[draw,shape=rectangle,inner sep=2pt,minimum height=6mm,minimum width=6mm,fill=white]
\tikzstyle{dbox}=[draw,doubled,shape=rectangle,inner sep=2pt,minimum height=6mm,minimum width=6mm,fill=white]
\tikzstyle{dmap}=[draw,doubled,shape=NEbox,inner sep=2pt,minimum height=6mm,fill=white]
\tikzstyle{dmapdag}=[draw,doubled,shape=SEbox,inner sep=2pt,minimum height=6mm,fill=white]
\tikzstyle{dmapadj}=[draw,doubled,shape=SEbox,inner sep=2pt,minimum height=6mm,fill=white]
\tikzstyle{dmaptrans}=[draw,doubled,shape=SWbox,inner sep=2pt,minimum height=6mm,fill=white]
\tikzstyle{dmapconj}=[draw,doubled,shape=NWbox,inner sep=2pt,minimum height=6mm,fill=white]

\tikzstyle{ddmap}=[draw,doubled,dashed,shape=NEbox,inner sep=2pt,minimum height=6mm,fill=white]
\tikzstyle{ddmapdag}=[draw,doubled,dashed,shape=SEbox,inner sep=2pt,minimum height=6mm,fill=white]
\tikzstyle{ddmapadj}=[draw,doubled,dashed,shape=SEbox,inner sep=2pt,minimum height=6mm,fill=white]
\tikzstyle{ddmaptrans}=[draw,doubled,dashed,shape=SWbox,inner sep=2pt,minimum height=6mm,fill=white]
\tikzstyle{ddmapconj}=[draw,doubled,dashed,shape=NWbox,inner sep=2pt,minimum height=6mm,fill=white]

\boxshape{sNEbox}{0pt}{3pt}
\boxshape{sSEbox}{0pt}{-3pt}
\boxshape{sNWbox}{3pt}{0pt}
\boxshape{sSWbox}{-3pt}{0pt}
\tikzstyle{smap}=[draw,shape=sNEbox,fill=white]
\tikzstyle{smapdag}=[draw,shape=sSEbox,fill=white]
\tikzstyle{smapadj}=[draw,shape=sSEbox,fill=white]
\tikzstyle{smaptrans}=[draw,shape=sSWbox,fill=white]
\tikzstyle{smapconj}=[draw,shape=sNWbox,fill=white]

\tikzstyle{dsmap}=[draw,dashed,shape=sNEbox,fill=white]
\tikzstyle{dsmapdag}=[draw,dashed,shape=sSEbox,fill=white]
\tikzstyle{dsmaptrans}=[draw,dashed,shape=sSWbox,fill=white]
\tikzstyle{dsmapconj}=[draw,dashed,shape=sNWbox,fill=white]

\boxshape{mNEbox}{0pt}{10pt}
\boxshape{mSEbox}{0pt}{-10pt}
\boxshape{mNWbox}{10pt}{0pt}
\boxshape{mSWbox}{-10pt}{0pt}
\tikzstyle{mmap}=[draw,shape=mNEbox]
\tikzstyle{mmapdag}=[draw,shape=mSEbox]
\tikzstyle{mmaptrans}=[draw,shape=mSWbox]
\tikzstyle{mmapconj}=[draw,shape=mNWbox]

\tikzstyle{mmapgray}=[draw,fill=gray!40!white,shape=mNEbox]
\tikzstyle{smapgray}=[draw,fill=gray!40!white,shape=sNEbox]

\makeatletter

\pgfdeclareshape{cornerpoint}{
\inheritsavedanchors[from=rectangle] 
\inheritanchorborder[from=rectangle]
\inheritanchor[from=rectangle]{center}
\inheritanchor[from=rectangle]{north}
\inheritanchor[from=rectangle]{south}
\inheritanchor[from=rectangle]{west}
\inheritanchor[from=rectangle]{east}
\backgroundpath{
\southwest \pgf@xa=\pgf@x \pgf@ya=\pgf@y
\northeast \pgf@xb=\pgf@x \pgf@yb=\pgf@y

\pgfmathsetmacro{\pgf@shorten@left}{\pgfkeysvalueof{/tikz/shorten left}}
\pgfmathsetmacro{\pgf@shorten@right}{\pgfkeysvalueof{/tikz/shorten right}}

\pgfpathmoveto{\pgfpoint{0.5 * (\pgf@xa + \pgf@xb)}{\pgf@ya - 5pt}}
\pgfpathlineto{\pgfpoint{\pgf@xa - 8pt + \pgf@shorten@left}{\pgf@yb - 1.5 * \pgf@shorten@left}}
\pgfpathlineto{\pgfpoint{\pgf@xa - 8pt + \pgf@shorten@left}{\pgf@yb}}
\pgfpathlineto{\pgfpoint{\pgf@xb + 8pt - \pgf@shorten@right}{\pgf@yb}}
\pgfpathlineto{\pgfpoint{\pgf@xb + 8pt - \pgf@shorten@right}{\pgf@yb - 1.5 * \pgf@shorten@right}}
\pgfpathclose
}
}

\pgfdeclareshape{cornercopoint}{
\inheritsavedanchors[from=rectangle] 
\inheritanchorborder[from=rectangle]
\inheritanchor[from=rectangle]{center}
\inheritanchor[from=rectangle]{north}
\inheritanchor[from=rectangle]{south}
\inheritanchor[from=rectangle]{west}
\inheritanchor[from=rectangle]{east}
\backgroundpath{
\southwest \pgf@xa=\pgf@x \pgf@ya=\pgf@y
\northeast \pgf@xb=\pgf@x \pgf@yb=\pgf@y

\pgfmathsetmacro{\pgf@shorten@left}{\pgfkeysvalueof{/tikz/shorten left}}
\pgfmathsetmacro{\pgf@shorten@right}{\pgfkeysvalueof{/tikz/shorten right}}

\pgfpathmoveto{\pgfpoint{0.5 * (\pgf@xa + \pgf@xb)}{\pgf@yb + 5pt}}
\pgfpathlineto{\pgfpoint{\pgf@xa - 8pt + \pgf@shorten@left}{\pgf@ya + 1.5 * \pgf@shorten@left}}
\pgfpathlineto{\pgfpoint{\pgf@xa - 8pt + \pgf@shorten@left}{\pgf@ya}}
\pgfpathlineto{\pgfpoint{\pgf@xb + 8pt - \pgf@shorten@right}{\pgf@ya}}
\pgfpathlineto{\pgfpoint{\pgf@xb + 8pt - \pgf@shorten@right}{\pgf@ya + 1.5 * \pgf@shorten@right}}
\pgfpathclose
}
}

\makeatother

\pgfkeyssetvalue{/tikz/shorten left}{0pt}
\pgfkeyssetvalue{/tikz/shorten right}{0pt}

\tikzstyle{kpoint common}=[draw,fill=white,inner sep=1pt,minimum height=4mm]
\tikzstyle{kpoint sc}=[shape=cornerpoint,kpoint common]
\tikzstyle{kpoint adjoint sc}=[shape=cornercopoint,kpoint common]
\tikzstyle{kpoint}=[shape=cornerpoint,shorten left=5pt,kpoint common]
\tikzstyle{kpoint adjoint}=[shape=cornercopoint,shorten left=5pt,kpoint common]
\tikzstyle{kpoint conjugate}=[shape=cornerpoint,shorten right=5pt,kpoint common]
\tikzstyle{kpoint transpose}=[shape=cornercopoint,shorten right=5pt,kpoint common]
\tikzstyle{kpoint symm}=[shape=cornerpoint,shorten left=5pt,shorten right=5pt,kpoint common]

\tikzstyle{black kpoint}=[shape=cornerpoint,shorten left=5pt,kpoint common,fill=black,font=\color{white}]
\tikzstyle{black kpoint adjoint}=[shape=cornercopoint,shorten left=5pt,kpoint common,fill=black,font=\color{white}]
\tikzstyle{black kpointadj}=[shape=cornercopoint,shorten left=5pt,kpoint common,fill=black,font=\color{white}]

\tikzstyle{black dkpoint}=[shape=cornerpoint,shorten left=5pt,kpoint common,fill=black, doubled,font=\color{white}]
\tikzstyle{black dkpoint adjoint}=[shape=cornercopoint,shorten left=5pt,kpoint common,fill=black, doubled,font=\color{white}]
\tikzstyle{black dkpointadj}=[shape=cornercopoint,shorten left=5pt,kpoint common,fill=black, doubled,font=\color{white}] 

\tikzstyle{kpointdag}=[kpoint adjoint]
\tikzstyle{kpointadj}=[kpoint adjoint]
\tikzstyle{kpointconj}=[kpoint conjugate]
\tikzstyle{kpointtrans}=[kpoint transpose]

\tikzstyle{big kpoint}=[kpoint, minimum width=1.2 cm, minimum height=8mm, inner sep=4pt, text depth=3mm]

\tikzstyle{wide kpoint}=[kpoint, minimum width=1 cm, inner sep=2pt]
\tikzstyle{wide kpointdag}=[kpointdag, minimum width=1 cm, inner sep=2pt]
\tikzstyle{wide kpointconj}=[kpointconj, minimum width=1 cm, inner sep=2pt]
\tikzstyle{wide kpointtrans}=[kpointtrans, minimum width=1 cm, inner sep=2pt]

\tikzstyle{gray kpoint}=[kpoint,fill=gray!50!white]
\tikzstyle{gray kpointdag}=[kpointdag,fill=gray!50!white]
\tikzstyle{gray kpointadj}=[kpointadj,fill=gray!50!white]
\tikzstyle{gray kpointconj}=[kpointconj,fill=gray!50!white]
\tikzstyle{gray kpointtrans}=[kpointtrans,fill=gray!50!white]

\tikzstyle{gray dkpoint}=[kpoint,fill=gray!50!white,doubled]
\tikzstyle{gray dkpointdag}=[kpointdag,fill=gray!50!white,doubled]
\tikzstyle{gray dkpointadj}=[kpointadj,fill=gray!50!white,doubled]
\tikzstyle{gray dkpointconj}=[kpointconj,fill=gray!50!white,doubled]
\tikzstyle{gray dkpointtrans}=[kpointtrans,fill=gray!50!white,doubled]

\tikzstyle{white label}=[draw,fill=white,rectangle,inner sep=0.7 mm]
\tikzstyle{gray label}=[draw,fill=gray!50!white,rectangle,inner sep=0.7 mm]
\tikzstyle{black label}=[draw,fill=black,rectangle,inner sep=0.7 mm]

\tikzstyle{dkpoint}=[kpoint,doubled]
\tikzstyle{wide dkpoint}=[wide kpoint,doubled]
\tikzstyle{dkpointdag}=[kpoint adjoint,doubled]
\tikzstyle{wide dkpointdag}=[wide kpointdag,doubled]
\tikzstyle{dkcopoint}=[kpoint adjoint,doubled]
\tikzstyle{dkpointadj}=[kpoint adjoint,doubled]
\tikzstyle{dkpointconj}=[kpoint conjugate,doubled]
\tikzstyle{dkpointtrans}=[kpoint transpose,doubled]

\tikzstyle{kscalar}=[kpoint common, shape=EBox, inner xsep=-1pt, inner ysep=3pt,font=\small]
\tikzstyle{kscalarconj}=[kpoint common, shape=WBox, inner xsep=-1pt, inner ysep=3pt,font=\small]

\tikzstyle{spekpoint}=[kpoint sc,minimum height=5mm,inner sep=3pt]
\tikzstyle{spekcopoint}=[kpoint adjoint sc,minimum height=5mm,inner sep=3pt]

\tikzstyle{dspekpoint}=[spekpoint,doubled]
\tikzstyle{dspekcopoint}=[spekcopoint,doubled]

 \tikzstyle{discard}=[ground,rotate=180,scale=1.5,inner sep=-2mm]
 \tikzstyle{downground}=[circuit ee IEC,thick,ground,rotate=-90,scale=1.5,inner sep=-2mm]

\tikzstyle{maxmix}=[regular polygon,regular polygon sides=3,draw=black,xscale=0.4,yscale=0.3,inner sep=-0.5pt,minimum width=10mm,fill=gray,regular polygon rotate=180]

 \tikzstyle{bigground}=[regular polygon,regular polygon sides=3,draw=gray,scale=0.50,inner sep=-0.5pt,minimum width=10mm,fill=gray]

\tikzstyle{arrs}=[-latex,font=\small,auto]
\tikzstyle{arrow plain}=[arrs]
\tikzstyle{arrow dashed}=[dashed,arrs]
\tikzstyle{arrow bold}=[very thick,arrs]
\tikzstyle{arrow hide}=[draw=white!0,-]
\tikzstyle{arrow reverse}=[latex-]
\tikzstyle{cdnode}=[]

\tikzstyle{green dashed arrow}=[green, arrow dashed]
\tikzstyle{dashed blue}=[blue, dashed]
\tikzstyle{red dashed arrow}=[red, arrow dashed]
\tikzstyle{orange arrow}=[orange, arrs]
\tikzstyle{blue arrow}=[blue, arrs]
\tikzstyle{magenta arrow}=[magenta, arrs]

\tikzstyle{dotted line}=[-, style=dotted, tikzit draw=brown]
\tikzstyle{dashed line}=[-, style=dashed, tikzit draw=cyan]
\tikzstyle{green fill line}=[-, fill={green!90!black}, tikzit draw=green]
\tikzstyle{blue fill}=[-, fill=blue, tikzit fill=blue, tikzit draw={rgb,255: red,102; green,117; blue,255}]
\tikzstyle{red}=[-, draw=red, tikzit draw=red]
\tikzstyle{blue}=[-, draw=blue, tikzit draw=blue]
\tikzstyle{thick black}=[-, draw=black, tikzit draw=black, line width=1pt]
\tikzstyle{dotted red}=[-, draw=red, style=dotted, tikzit draw=red]
\tikzstyle{dotted blue}=[-, draw=blue, tikzit draw=blue, style=dotted]
\tikzstyle{dashed thick blue}=[-, draw={rgb,255: red,28; green,176; blue,255}, tikzit draw={rgb,255: red,83; green,19; blue,156}, line width=1pt, style=dashed]
\tikzstyle{dashed thick red}=[-, draw=red, tikzit draw={rgb,255: red,255; green,100; blue,10}, line width=1pt, style=dashed]
\tikzstyle{green}=[-, draw=green, tikzit draw=green]
\tikzstyle{dotted green}=[-, draw=green, tikzit draw=green, style=dotted]
\tikzstyle{arrow}=[->]
\tikzstyle{arrow green dashed}=[draw=green, ->, tikzit draw=green, style=dashed]
\tikzstyle{arrow dashed red}=[draw=red, ->, style=dashed, tikzit draw=red]
\tikzstyle{dashed green}=[-, tikzit draw=green, draw=green, style=dashed]

\DeclareFontSeriesDefault[rm]{bf}{b}

\usepackage{abstract}
\AtBeginDocument{} 

\usepackage[font=small,labelfont=it]{caption}

\usepackage{tikz}
\usepackage{lipsum}

\usepackage{amsthm, bm}
\usepackage{mdframed}

\theoremstyle{definition}
\newtheorem{example}{Example}[section]

\newtheorem{remark}{Remark}[section]
\theoremstyle{plain}

\newenvironment{examplebox}
  {\begin{mdframed}[
     linewidth=0.8pt,
     roundcorner=5pt,
     innertopmargin=1pt,
     innerbottommargin=12pt,
     innerleftmargin=8pt,
     innerrightmargin=8pt
   ]
   \begin{example}}
  {\end{example}
   \end{mdframed}}

\newcommand{\gen}[1]{\left\langle #1 \right\rangle}

\newcommand{\cc}{\mathcal C}

\newcommand{\ch}{\mathcal H}
\newcommand{\ci}{\mathcal I}

\newcommand{\cl}{\mathcal L}

\newcommand{\cu}{\mathcal U}

\newcommand{\cx}{\mathcal X}

\newcommand{\bA}{{\bm A}}
\newcommand{\bB}{{\bm B}}
\newcommand{\bC}{{\bm C}}
\newcommand{\bD}{{\bm D}}
\newcommand{\bE}{{\bm E}}
\newcommand{\bF}{{\bm F}}
\newcommand{\bG}{{\bm G}}

\newcommand{\bJ}{{\bm J}}

\newcommand{\bM}{{\bm M}}

\newcommand{\bO}{{\bm O}}
\newcommand{\bP}{{\bm P}}
\newcommand{\bQ}{{\bm Q}}

\newcommand{\bS}{{\bm S}}
\newcommand{\bT}{{\bm T}}

\newcommand{\bX}{{\bm X}}
\newcommand{\bY}{{\bm Y}}
\newcommand{\bZ}{{\bm Z}}

\usepackage{bettercenternot}

\usepackage[normalem]{ulem}

\usepackage{authblk}

\title{\Huge \bf Ontology without states: \\ [0.5ex] \Large  PBR as an argument for dynamics-first realism}
\author[1]{Nick Ormrod\footnote{normrod@perimeterinstitute.ca}}
\affil[1]{\small Perimeter Institute for Theoretical Physics,  31 Caroline Street North, Waterloo, Ontario, N2L 2Y5, Canada}
\date{\today}

\begin{document}
    \maketitle

\begin{abstract}
The Pusey--Barrett--Rudolph (PBR) theorem is commonly taken to establish that a pure quantum state represents a physical property of the system to which it is assigned, rather than merely encoding knowledge. 
In this work, I distinguish between state-first and dynamics-first physical theories, and argue that this reading presupposes the former. These two types of theories support different approaches to causal explanation, and one of PBR's stated assumptions depends on the state-first approach. 
Using a recent interpretation of quantum theory as a concrete example, I show how a dynamics-first theory can explain the PBR correlations while maintaining a broadly epistemic reading of quantum states. 
I conclude that the PBR theorem supports a dilemma: either the quantum state is part of reality, or reality is not fundamentally built out of states. 
\end{abstract}

\tableofcontents

\section{Introduction}
A debate of central importance for the foundations of quantum theory concerns the role of a pure quantum state: is it part of reality, or merely a description of an agent's knowledge? It is often suggested that the debate has been settled in favor of the former by the Pusey-Barrett-Rudolph (PBR) theorem \citep{pusey2012reality}. But this paper will introduce a distinction between \textit{state-first} and \textit{dynamics-first} physical theories, and show that it is only the former type of theory that struggles to explain the PBR correlations without granting reality to the quantum state.

The distinction between state-first and dynamics-first theories concerns the conceptual relationship between states and dynamics. 
In a state-first theory, dynamics supervene\footnote{A variable $B$ \textit{supervenes} on a variable $A$ precisely if the value of $A$ uniquely determines the value of $B$, i.e.\ if $B$ can be calculated via some function $f: A \rightarrow B$.} on relations of counterfactual dependence between states. 
History is awash with such theories. 
Their laws are often expressed by equations showing how the states of different subsystems evolve over time, supporting counterfactual propositions about how the state of this system at this time would have been different had the state of that system at that time taken a different value. 
Once one knows everything about how the dynamics constrain the relationship between different systems' states over time, one understands the dynamics fully: there is nothing else that they do. 

Dynamics-first theories are different. 
One essential feature of a dynamics-first theory is that dynamics are \textit{primitive}, meaning that they do not supervene on counterfactual dependence relations between states. 
These relations supervene on the dynamics, but the dynamics have structure over and above them.\footnote{As we will clarify in \cref{ft:leibniz}, it does not follow that dynamics-first theories have structure in their ontologies that lack operational consequence.}
The other essential feature of a dynamics-first theory is that states emerge from dynamical interactions, and only from interactions that have a structure that allows them to do so. 
This means that when the relevant interactions do not have such a structure, the interacting systems do not have states.

If it sounds incomprehensible that systems could interact without having any states, then state-first presuppositions are to blame. 
On the state-first view, the idea of dynamical interactions between stateless systems is indeed nonsensical: if dynamics supervene on counterfactual dependence relations between states, then there can obviously be no dynamical interactions without states. But in a dynamics-first theory, the dynamics are primitive, the supervenience claim fails, and thus a system's lack of a state does not preclude it from participating in the dynamics. 


History has not been a generous provider of dynamics-first theories, but a recent interpretation of quantum theory provides one concrete example of what one might look like. In \textit{Dynamics-First Quantum Theory}\footnote{Dynamics-First Quantum Theory was called \textit{Causal Decoherence Theory} in \citep{ormrod2026decoherence}.} \citep{ormrod2024quantum, ormrod2026decoherence} (DFQT), the unitary dynamics are taken as a primitive entity, and not as a rule governing the evolution of any state. Instead, all physical states in the theory are held to emerge from the unitary dynamics, and only when they have a decohering structure. The theory describes an observer-independent reality through a precise mathematical formalism which reproduces the quantum predictions---thus demonstrating that the dynamics-first philosophy can be realized in a physical theory that saves the phenomena.

PBR's argument for realism about pure quantum states relies on an approach to causal explanation that is supported by state-first theories but undermined by dynamics-first ones. Indeed, consider a \enquote{prepare then measure} in which there is an observed correlation between setting $S$ on the device that prepares the system and the outcome $O$ of the subsequent measurement, i.e.\ ${\rm Prob}(O|S) \neq {\rm Prob}(O)$. PBR's \textit{screening-off} assumption requires that this correlation is explained by the physical state $\Lambda$ of the prepared system, and that the explanation works by showing that the correlation is eliminated by conditioning on the physical state $\Lambda$ of the prepared system, i.e.\ 
\begin{equation} \label{eq:screening_off}
{\rm Prob}(O|\Lambda, S) = {\rm Prob}(O|\Lambda).
\end{equation}

From the state-first point of view, the screening-off assumption is well-motivated. According to any state-first theory, $S$ can influence $\Lambda$ only insofar as counterfactual variations in $S$ induce counterfactual variations in $\Lambda$. Hence, if the dynamical connection between $S$ and $O$ consists only of an influence that is fully mediated by $\Lambda$, then no correlation between $S$ and $O$ should remain once we condition on a fixed state $\Lambda=\lambda$. Thus state-first ontology, together with this plausible assumption about the causal structure of the experiment, entails screening-off. However, from the dynamics-first point of view, the screening-off assumption is not well-motivated.
In fact, it is undermined: in a dynamics-first theory, a system can mediate an influence without having a state at all, and thus there should be no expectation that it has a state satisfying the screening-off assumption.

Dynamics-first theories thus require an alternative approach to causal explanation. 
I propose such an alternative: a correlation is explained by a given causal connection when the correlation would not survive a minimal counterfactual modification to the dynamics that breaks that connection. 
In other words, causal explanations are given by using the theory to demonstrate that in a nearby possible world in which the relevant causal connection is absent, the correlation does not arise---thus showing that in the actual world, the causal connection is responsible for the correlation.
In the case of a \enquote{prepare then measure} experiment, this means showing that if the causal connection between the preparation and measurement devices, bridged by the prepared system, had been broken, then there would have been no correlation between the preparation setting and the measurement outcome. This approach to causal explanation does not require that the prepared system has any physical state, much less one that screens off the correlation.

This work will show that on this dynamics-first approach to causal explanation, the PBR correlations can be comfortably explained while maintaining a broadly epistemic reading of quantum states. Throughout the argument, DFQT serves as a concrete proof-of-principle: we will use this theory to build a precise model of the PBR scenario which reproduces the quantum predictions exactly, and does so without reifying the quantum state. We will also demonstrate that it provides a causal explanation of those predictions in the sense just proposed.

For anyone wedded to the state-first approach to ontology, the current work will not change the moral of the PBR theorem: given such a commitment, the screening-off approach to causal explanation remains natural, and the theorem remains a powerful argument for regarding the pure quantum state of a system as a property of its physical state.
But if one's prior commitment is instead to an ontology free of quantum states, PBR turns out to be a powerful argument for putting the dynamics first.

This paper is structured as follows. \cref{sec:pbr} introduces the PBR theorem, and can be skipped by readers who are already confident with the PBR theorem and the precise version of quantum state realism (dubbed $\psi$-ontology) defined in \citep{harrigan2010einstein}. \cref{sec:cdt} introduces DFQT and provides a DFQT model of the PBR scenario. \cref{sec:reconciliation} shows how the PBR correlations can be reconciled with anti-realism about the quantum state, and \cref{sec:explanation} shows how these correlations are explained. \cref{sec:epistemic} discusses the sense in which DFQT supports an \enquote{epistemic} reading of quantum states before \cref{sec:conc} concludes the work.

\section{The PBR theorem} \label{sec:pbr} 

What does it mean to assert that a variable $V$ is a property of a system? A plausible answer is that $V$ supervenes on the complete physical state $\Lambda$ of the system. Hence the energy of a classical particle is a property of the system, because it depends only on where that particle is in its phase space. But the probability distribution over possible energies assigned by an agent is not a property of the system: the phase-space location of the particle does not determine the probability distribution, because the distribution depends on other factors, such as the information available to the agent. 

Accepting this definition of a property, the precise notion of quantum state realism relevant to PBR is that the pure quantum state assigned to a system is a property of that system: unlike a probability distribution, it depends on nothing more than the system's physical state.
The claim is not that the quantum state is a complete description of the physical state; completeness would mean that the physical state $\Lambda$ is determined by the quantum state $\Psi$ via a function $g:\Psi \rightarrow \Lambda$, whereas realism means only that $\Psi$ is determined via a function $f:\Lambda \rightarrow \Psi$.
Thus Bohm theory, which claims that the physical state of a system consists of both the quantum state and a configuration of particles, asserts that the quantum state is real, though incomplete. This notion of quantum state realism was introduced in \citep{harrigan2010einstein}, and is often called \textit{$\psi$-ontology}. Its denial can be called $\psi$\textit{-epistemology}. 

The PBR theorem shows that, given some assumptions, $\psi$-ontology is true. However, for our purposes, it is not necessary to reconstruct the full PBR argument. 
It will suffice to recount a special case of the argument that seeks to establish a corollary of $\psi$-ontology. This corollary concerns a situation in which a quantum system is prepared in one of two quantum states, $\ket {\psi_0}$ or $\ket{\psi_1}$, with an inner product of $\braket{\psi_0|\psi_1} = \frac{1}{\sqrt{2}}$. It says that the quantum state of the qubit in this situation is determined by its physical state $\Lambda$ via a function
\begin{equation} \label{eq:restricted_psi_ontic}
    f: \qquad \Lambda \quad \rightarrow \quad \{{ \ket{\psi_0}, \ket{\psi_1}}\}.
\end{equation}

The special PBR argument considers a scenario in which two qubits, $\bA$ and $\bB$, are each prepared in either $\ket{\psi_0}$ or $\ket{\psi_1}$ and then jointly measured in an entangled basis. 
To define this measurement basis, we first note that we can always find a pair of local orthonormal bases, $\{\ket 0 , \ket{1}\}$ and $\{\ket +, \ket -\}$, for each qubit such that $\ket{\psi_0} = \ket 0$, $\ket {\psi_1} = \ket +$, and $\ket{+} = \frac{1}{\sqrt{2}}(\ket 0 + \ket 1)$. The measurement basis is then given by
\begin{equation}\label{eq:q_meas}
\begin{aligned}
\ket{\xi_{00}} &:= \frac{1}{\sqrt{2}}\bigl(\ket{0}_\bA\ket{1}_\bB + \ket{1}_\bA\ket{0}_\bB\bigr),
&\qquad
\ket{\xi_{01}} &:= \frac{1}{\sqrt{2}}\bigl(\ket{0}_\bA\ket{-}_\bB + \ket{1}_\bA\ket{+}_\bB\bigr),
\\
\ket{\xi_{10}} &:= \frac{1}{\sqrt{2}}\bigl(\ket{+}_\bA\ket{1}_\bB + \ket{-}_\bA\ket{0}_\bB\bigr),
&\qquad
\ket{\xi_{11}} &:= \frac{1}{\sqrt{2}}\bigl(\ket{+}_\bA\ket{-}_\bB + \ket{-}_\bA\ket{+}_\bB\bigr).
\end{aligned}
\end{equation}

In this scenario, quantum theory predicts a conditional probability distribution ${\rm Prob}(O|S,T)$ for the outcome $O$ of the measurement given the settings $S, T$ on the devices that prepare $\bA$ and $\bB$. Because  $\braket{\xi_{ij} | \psi_i,\psi_j} =0$ for all $i, j \in \{0, 1\}$, the predictions entail that whenever the outcome associated with $\ket{\xi_{ij}}$ is obtained, the initial state $\ket{\psi_i,\psi_j}$ is ruled out with certainty, i.e.\ 
\begin{equation} \label{eq:zero_probs}
{\rm Prob}(O=ij|S=i,T=j)=0
\end{equation}
for all $i, j \in \{0, 1\}$. This is the key property of the observed statistics to be exploited by the argument.

The space of complete physical states of the composite system $\bA\bB$ is denoted by $\Lambda_{\bA\bB}$. To avoid the need for measure-theoretic notation, we will assume that $\Lambda_{\bA\bB}$ is discrete. The argument can be straightforwardly reformulated for general state spaces in terms of supports of probability distributions. 

The first substantive assumption of the argument is \textit{screening off}, expressed for a general \enquote{prepare then measure} scenario by \cref{eq:screening_off}. In the PBR scenario, this implies that
\begin{equation} \label{eq:screening_off_2} {\rm Prob}(O| S,T)=\sum_{\lambda_{\bA\bB}\in\Lambda_{\bA\bB}} {\rm Prob}(\Lambda_{\bA\bB}=\lambda_{\bA\bB}| S,T)\ {\rm Prob}(O|\Lambda_{\bA\bB}=\lambda_{\bA\bB}). \end{equation}
Since each term in \cref{eq:screening_off_2} is non-negative, it follows by \cref{eq:zero_probs} that any state $\lambda_{\bA\bB}$ compatible with a given choice $S=i$, $T=j$ of settings must attribute zero probability to the corresponding outcome $O=ij$. That is, for all $i, j \in \{0, 1\}$ and $\lambda_{\bA\bB} \in \Lambda_{\bA\bB}$, if ${\rm Prob}(\Lambda_{\bA\bB} = \lambda_{\bA\bB}| S=i,T=j)>0$ then $
{\rm Prob}(O=ij|\Lambda_{\bA\bB}=\lambda_{\bA\bB})=0$. Since every state must attribute a nonzero probability to at least one possible outcome, it follows that there exists no physical state of the pair compatible with all four pairs of preparation settings:
\begin{equation} \label{eq:four_way_comp}
\nexists\,\lambda_{\bA\bB}\in\Lambda_{\bA\bB}
\quad\text{such that}\quad
{\rm Prob}(\Lambda_{\bA\bB}=\lambda_{\bA\bB}\mid S=i,T=j)>0
\quad\text{for all } i,j\in\{0,1\}.
\end{equation}

We now require some further assumptions. \textit{Ontological separability} asserts that the physical state of the composite system $\bA\bB$ is specified by a pair $(\lambda_{\bA},\lambda_{\bB})\in\Lambda_{\bA\bB}$ of physical states of the individual systems. In case ontological separability seems strong, note that it need not hold generically, but only for systems that are prepared by independent procedures. For independent preparation procedures, one also naturally expects that the conditional probability distribution factorizes,
\begin{equation} \label{eq:prep_ind} {\rm Prob}(\lambda_{\bA},\lambda_{\bB} | S,T)={\rm Prob}(\lambda_{\bA} |S)\,{\rm Prob}(\lambda_{\bB} |T). 
\end{equation}
This assumption is called \textit{preparation independence}.
Finally, assuming that the preparation devices for $\bA$ and $\bB$ have been designed in the same way, it is fair to assume that they give rise to conditional probability distributions for physical states given settings that are identical up to relabelling of the distributed variables; we denote this condition by
\begin{equation} \label{eq:probs_identical}
    {\rm Prob}(\lambda_{\bA}| S) \cong {\rm Prob}(\lambda_{\bB}| T).
\end{equation}

Now suppose, for contradiction, that the quantum state assigned to $\bA$ was not determined via a function $f: \Lambda_\bA \rightarrow \{\ket{\psi_0}_\bA, \ket{\psi_1}_\bA\}$. This would mean that there exists a single physical state $\lambda_{\bA}$ compatible with both $\ket{\psi_0}_\bA$ and $\ket{\psi_1}_\bA$, i.e.\ ${\rm Prob}(\lambda_{\bA} |  S=i)>0$ for both $i\in \{0, 1\}$. By \cref{eq:probs_identical}, it follows that there is an analogous physical state $\lambda_{\bB}$ of $\bB$, satisfying ${\rm Prob}(\lambda_{\bB} |  T=j)>0$ for both $j\in \{0, 1\}$. By ontological separability, these two states define a joint state $\lambda_{\bA\bB}=(\lambda_{\bA},\lambda_{\bB})$ of $\bA\bB$, and, by the preparation independence assumption of \cref{eq:prep_ind}, this joint state is a counterexample to \cref{eq:four_way_comp}, giving us our contradiction.

Given the assumptions, we have therefore established that the corollary of $\psi$-ontology for the qubit $\bA$ in this scenario is true: the quantum state $\ket{\psi_i}_\bA$ assigned to the qubit depends only on the physical state of the qubit, and is thus determined via a function $f: \Lambda_\bA \rightarrow \{{ \ket{\psi_0}_\bA, \ket{\psi_1}_\bA}\}$.
(By a similar argument, the analogous statement also holds for $\bB$.)

The screening-off assumption played an essential role in the PBR argument, allowing the derivation of \cref{eq:four_way_comp}. And, as we already saw in the introduction, this assumption is undermined by the dynamics-first conception of ontology, leaving open the possibility of a dynamics-first explanation of the PBR correlations that escapes the $\psi$-ontic conclusion.
But that discussion was quite abstract---we are still yet to see a proof of principle demonstrating that dynamics-first theories are actually possible, much less that one can explain the PBR correlations without granting reality to the quantum state.
Our next task, then, is to introduce the dynamics-first interpretation of quantum theory recently developed in \citep{ormrod2024quantum, ormrod2026decoherence}.

\section{Dynamics-First Quantum Theory} \label{sec:cdt}

In quantum theory, the unitary dynamics are typically regarded as a rule governing the evolution of a quantum state. 
But Dynamics-First Quantum Theory denies this, holding instead that the unitary dynamics are a fundamental feature of the ontology: they “just exist”. 

When unitary interactions have a decohering structure, events emerge from them. But when none of the interactions have a decohering structure, nothing exists except the interactions themselves. Hence DFQT has both essential features of a dynamics-first theory: it takes dynamics as primitive, and it holds that physical states (which in DFQT are constituted by events) emerge from dynamical interactions that have a certain structure.

This section will provide an introduction to DFQT.
Compared to existing expositions in \citep{ormrod2024quantum, ormrod2026decoherence}, the one here will be more streamlined and focus more closely on DFQT's status as a concrete example of a dynamics-first theory.
\cref{subsec:bipartite} and \cref{subsec:dem} will lay out the dynamical theory of quantum decoherence that underpins DFQT, before \cref{subsec:first} introduces the theory itself. Much of the discussion will proceed by developing a model of the PBR scenario, which in the current section serves as a pedagogic illustration of DFQT but later on will form the basis of the dynamics-first response to PBR.

\subsection{The dynamics of decoherence} \label{subsec:bipartite}

Discussion of decoherence often focuses on state-level features such as the suppression of off-diagonal terms in a density matrix. 
But DFQT is built on an approach to decoherence that puts the spotlight on the underlying dynamical structures, characterizing the dynamics of decoherence directly rather than diagnosing them through their state-level symptoms.

This approach to decoherence, developed in \citep{ormrod2026decoherence}, is best explained beginning with the simple case of a bipartite unitary interaction represented by a channel $\cu: \cl(\ch_\bS \otimes \ch_\bF) \rightarrow \cl(\ch_\bT \otimes \ch_\bG)$. For pedagogical purposes, we will think of this channel as an interaction between a \enquote{system} and a fragment of the \enquote{environment},
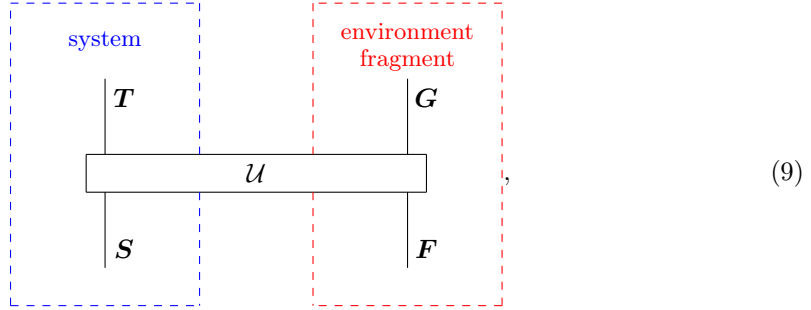
\begin{equation} \label{eq:interaction}
    \begin{tikzpicture}
	\begin{pgfonlayer}{nodelayer}
		\node [style=none] (0) at (-4.5, -0.5) {};
		\node [style=none] (1) at (4.5, -0.5) {};
		\node [style=none] (2) at (-4.5, 0.5) {};
		\node [style=none] (3) at (4.5, 0.5) {};
		\node [style=none] (4) at (0, 0) {$\cu$};
		\node [style=none] (5) at (-4, -2.5) {};
		\node [style=none] (6) at (-4, -0.5) {};
		\node [style=none] (7) at (-4, 0.5) {};
		\node [style=none] (8) at (-4, 2.5) {};
		\node [style=none] (9) at (4, -2.5) {};
		\node [style=none] (10) at (4, -0.5) {};
		\node [style=none] (11) at (4, 0.5) {};
		\node [style=none] (12) at (4, 2.5) {};
		\node [style=none] (13) at (-3.5, -2) {$\bS$};
		\node [style=none] (14) at (-3.5, 2) {$\bT$};
		\node [style=none] (15) at (4.5, -2) {$\bF$};
		\node [style=none] (16) at (4.5, 2) {$\bG$};
		\node [style=none] (17) at (-1.5, 0.5) {};
		\node [style=none] (18) at (-1.5, 4.5) {};
		\node [style=none] (19) at (-4, 3.5) {\small \color{blue} system};
		\node [style=none] (20) at (4, 3.75) {\small \color{red} environment};
		\node [style=none] (21) at (-1.5, -3.5) {};
		\node [style=none] (22) at (-1.5, -0.5) {};
		\node [style=none] (23) at (-6.5, 4.5) {};
		\node [style=none] (24) at (-6.5, -3.5) {};
		\node [style=none] (33) at (1.5, 0.5) {};
		\node [style=none] (34) at (1.5, 4.5) {};
		\node [style=none] (36) at (1.5, -3.5) {};
		\node [style=none] (37) at (1.5, -0.5) {};
		\node [style=none] (38) at (6.5, 4.5) {};
		\node [style=none] (39) at (6.5, -3.5) {};
		\node [style=none] (40) at (4, 3) {\small \color{red} fragment};
	\end{pgfonlayer}
	\begin{pgfonlayer}{edgelayer}
		\draw (3.center) to (1.center);
		\draw (1.center) to (0.center);
		\draw (0.center) to (2.center);
		\draw (2.center) to (3.center);
		\draw (6.center) to (5.center);
		\draw (8.center) to (7.center);
		\draw (10.center) to (9.center);
		\draw (12.center) to (11.center);
		\draw [dashed, blue] (18.center) to (17.center);
		\draw [dashed, blue] (22.center) to (21.center);
		\draw [dashed, blue] (24.center) to (21.center);
		\draw [dashed, blue] (24.center) to (23.center);
		\draw [dashed, blue] (23.center) to (18.center);
		\draw [dashed, red] (34.center) to (33.center);
		\draw [dashed, red] (37.center) to (36.center);
		\draw [dashed, red] (39.center) to (36.center);
		\draw [dashed, red] (39.center) to (38.center);
		\draw [dashed, red] (38.center) to (34.center);
	\end{pgfonlayer}
\end{tikzpicture},
\end{equation}
though we persist in using different labels $\bS / \bT$ for the system before and after the interaction, and likewise $\bF / \bG$ for the environment fragment. This affords us greater generality, as the dimension of $\bS$ need not equal the dimension of $\bT$. For simplicity, we assume throughout that all quantum systems are finite-dimensional.

To characterize the dynamics of decoherence, we want to define a set of decohered operators purely in terms of the unitary channel $\cu$ (including the given factorizations of its input and output space), assuming nothing about the quantum states of the relevant systems.
It is helpful to identify systems with their associated algebras\footnote{Throughout this work, an \textit{algebra} means a von Neumann algebra of operators; in finite dimensions, a set of operators on a Hilbert space that contains the identity and is closed under complex combinations, products, and adjoints. A set $\bA$ of operators is a von Neumann algebra precisely if it both contains its own adjoints and is equal to its double-commutant, $\bA'' = \bA$.} of operators, e.g.\ $\bS:= \cl(\ch_\bS) \otimes I_\bF$ and $\bG:= I_\bT \otimes\cl(\ch_\bG)$.
The \textit{decohered subalgebra} $\bS^\cu_{{\rm dec}\bG}$  of $\bS$ is then defined as the intersection
\begin{equation} \label{eq:dec}
    \begin{split}
       & \bS^\cu_{{\rm dec}\bG}  := \bS^\cu_{{\rm sur}\bG} \cap \bS^\cu_{{\rm rep}\bG},
    \end{split}
\end{equation}
of subalgebras of $\bS$ that respectively \textit{survive} and \textit{reproduce},\footnote{These algebras were respectively called \textit{potentially accessible} and \textit{accessible} in \citep{ormrod2025causal, ormrod2026decoherence}.} themselves defined by
\begin{equation} \label{eq:sur_rep}
    \bS^\cu_{{\rm sur}\bG} := \cu^{-1}(\bG)' \cap \bS \qquad {\rm and} \qquad 
    \bS^\cu_{{\rm rep}\bG} := (\cu^{-1}(\bG)' \cap \bS)' \cap \bS,
\end{equation}
where the \textit{commutant} $\cu^{-1}(\bG)'$ of $\cu^{-1}(\bG)$ is the subset of $\cl(\ch_\bS \otimes \ch_\bF)$ that commutes with every operator in $\cu^{-1}(\bG)$. 

Before explaining the biological terminology, we first note that for any $\cu$, $\bS^\cu_{{\rm dec}\bG}$ is commutative, because it is the intersection of two algebras that commute with each other. As a result, for any $\cu$, decoherence privileges a unique decomposition 
\begin{equation} \label{eq:hspace_decomp}
    \ch_\bS = \bigoplus_{s \in S}\ch_\bS^s
\end{equation}
of the Hilbert space of the system such that
\begin{equation}
    \bS^\cu_{{\rm dec}\bG} = {\rm span}_{\mathbb C}(\pi_\bS^S),
\end{equation}
where $S = \{1, \ldots, n\}$ is a finite set and $\pi_\bS^S := \{\pi_\bS^s|s\in S\}$ is the projective-valued measurement (PVM) whose elements project onto the subspaces in \cref{eq:hspace_decomp}. $\cu$ has a nontrivially decohering structure if $n\neq 1$, and a maximally decohering structure if $n$ is equal to the dimension of $\ch_\bS$.

\citep{ormrod2026decoherence} gives a detailed motivation for this definition of decoherence drawing on quantum Darwinism (see e.g.\ \citep{zurek2009quantum, zurek2025decoherence} and the references therein), which argues that the classical world emerges via the redundant proliferation of information about effectively classical observables. Summarizing crudely, the algebra $\bS^\cu_{{\rm rep}\bG}$ is called \enquote{reproducing} because changes to its observables necessarily bring about, via $\cu$, changes to observables in $\bG$, and in that sense $\bG$ obtains information about these observables. $\bS^\cu_{{\rm sur}\bG}$ is called \enquote{surviving} because its observables are the ones that remain local system observables throughout the interaction, and thus remain capable of reproducing in other fragments via future interactions with the system. Hence the decohered observables are precisely the ones that both survive and reproduce, thereby coming a step closer to proliferating throughout the environment.

Some examples (adapted from \citep{ormrod2026decoherence}) should help the reader develop the right intuition for decoherence. In \cref{ex:balance_a}, all system observables survive but no nontrivial observables reproduce, so the decohered algebra is trivial, consisting only of scalar multiples of the identity. In \cref{ex:balance_b}, all reproduce but no nontrivial observables survive, and the net result is the same: the decohered algebra is trivial. Only in \cref{ex:balance_c} are there nontrivial observables that manage to do both, winning the Darwinian competition by surviving \textit{and} reproducing, which is to say decohering. More generally, the fact that $\bS^\cu_{{\rm dec}\bG}$ is commutative means it can never be equal to the full $\bS$, meaning that operators must always compete with one another to decohere.

\begin{figure} [t!]
\centering
\begin{center} \small
\input{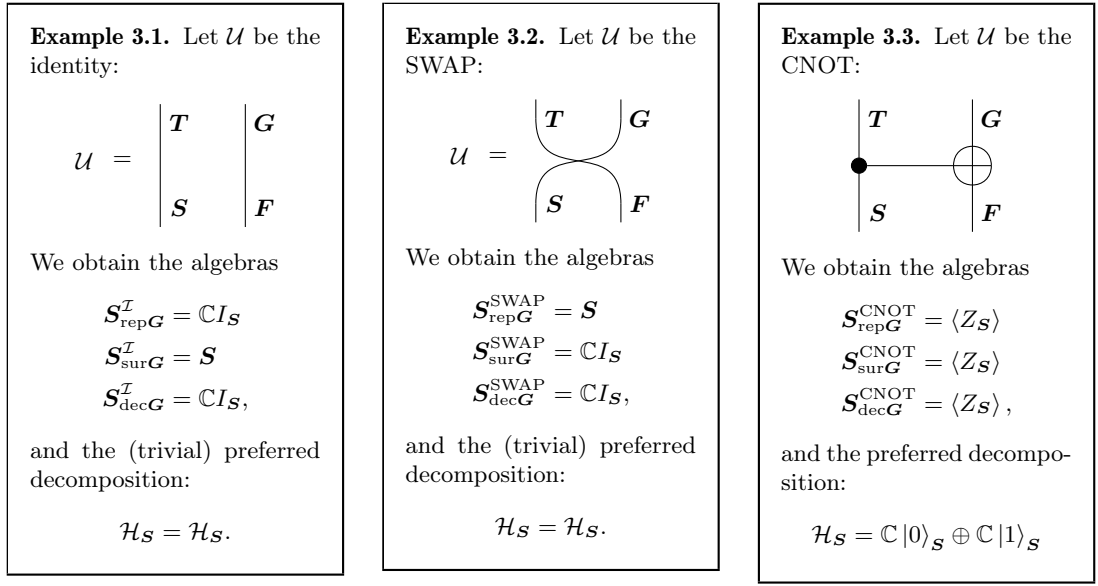}
\end{center}
    \caption{Three 2-qubit examples illustrating the dynamical notion of decoherence as survival and reproduction. $\gen{Z_\bS}$ is the algebra generated by the Pauli operator $Z_\bS$, i.e.\ the set of all system operators diagonal in the $Z_\bS$-basis. The CNOT unitary is defined at the Hilbert-space level by $\ket k_\bS \ket l_\bF \mapsto \ket k_\bT \ket{l+k}_\bG$, where \enquote{+} denotes addition modulo 2.
   }
    \label{fig:balance}
\end{figure}

While the analogy between quantum decoherence and Darwin's theory of natural selection is rewarding in its own right, our primary goal here will be to understand DFQT and its status as a dynamics-first theory.
With that goal in mind, the only essential thing to remember about the definition of the decohered algebra $\bS_{{\rm dec}\bG}^\cu$ is simply that it depends only on $\cu$; it does not depend on any assigned quantum state.
The purely dynamical status of $\bS_{{\rm dec}\bG}^\cu$ is crucial for the role it will come to play in putting the dynamics first.

\subsection{Democratic decoherence} \label{subsec:dem}

The previous subsection treated the \enquote{system} and the \enquote{environment} differently from one another: the former was what got decohered, and the latter was what did the decohering. 
But the analysis there can now be leveraged to show how systems in a given collection decohere \textit{each other}. 
This leads to a more democratic conception of decoherence in which each system in the collection serves as part of the environment that decoheres every other.

It is easiest to proceed with a concrete example, and, since our ultimate goal is to respond to the PBR theorem, that example may as well be the PBR scenario itself. The unitary circuit below is a (toy) model of the process by which the basis $\{\ket{\xi_{ij}}\}_{i, j=0}^1$ defined in \cref{eq:q_meas} is measured:
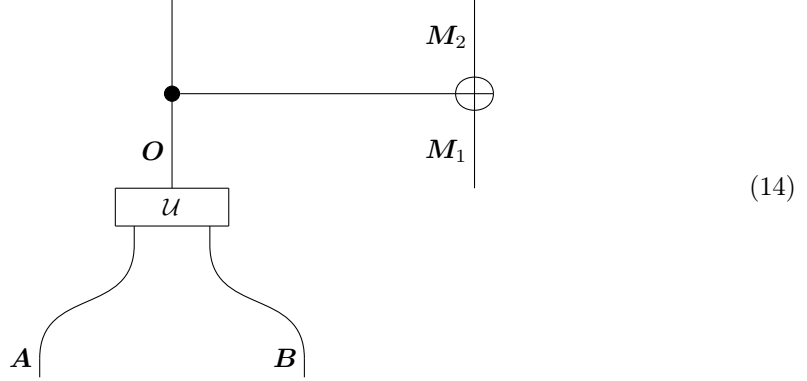
\begin{equation} \label{eq:mmt}
    \begin{tikzpicture}
	\begin{pgfonlayer}{nodelayer}
		\node [style=none] (0) at (1, -4.5) {};
		\node [style=none] (1) at (3.5, -1.5) {};
		\node [style=none] (2) at (12.5, 0) {};
		\node [style=none] (3) at (12.5, 5) {};
		\node [style=none] (4) at (3, -1) {};
		\node [style=none] (5) at (6, -1) {};
		\node [style=none] (6) at (6, 0) {};
		\node [style=none] (7) at (3, 0) {};
		\node [style=none] (8) at (4.5, 0) {};
		\node [style=none] (9) at (4.5, 5) {};
		\node [style=black dot] (10) at (4.5, 2.5) {};
		\node [style=none] (11) at (4.5, -0.5) {\small $\cu$};
		\node [style=none] (12) at (4, 1) {$\bO$};
		\node [style=none] (13) at (0.5, -4.5) {$\bA$};
		\node [style=none] (14) at (11.75, 4) {$\bM_2$};
		\node [style=none] (15) at (3.5, -1) {};
		\node [style=none] (16) at (1, -5) {};
		\node [style=none] (17) at (8, -4.5) {};
		\node [style=none] (18) at (5.5, -1.5) {};
		\node [style=none] (19) at (7.5, -4.5) {$\bB$};
		\node [style=none] (20) at (5.5, -1) {};
		\node [style=none] (21) at (8, -5) {};
		\node [style=none] (22) at (13, 2.5) {};
		\node [style=none] (23) at (12, 2.5) {};
		\node [style=none] (24) at (13, 2.5) {};
		\node [style=none] (25) at (12, 2.5) {};
		\node [style=none] (26) at (11.75, 1) {$\bM_1$};
	\end{pgfonlayer}
	\begin{pgfonlayer}{edgelayer}
		\draw [in=90, out=-90, looseness=1.25] (1.center) to (0.center);
		\draw (7.center) to (4.center);
		\draw (4.center) to (5.center);
		\draw (5.center) to (6.center);
		\draw (6.center) to (7.center);
		\draw (8.center) to (9.center);
		\draw (15.center) to (1.center);
		\draw (16.center) to (0.center);
		\draw [in=90, out=-90, looseness=1.25] (18.center) to (17.center);
		\draw (20.center) to (18.center);
		\draw (21.center) to (17.center);
		\draw (10) to (22.center);
		\draw (3.center) to (2.center);
		\draw [bend left=90, looseness=1.50] (23.center) to (22.center);
		\draw [bend right=90, looseness=1.50] (25.center) to (24.center);
	\end{pgfonlayer}
\end{tikzpicture}
\end{equation}
In this model of the measurement, the initially separated qubits $\bA$ and $\bB$ are first brought together. Then a unitary channel $\cu$ transforms the PBR measurement basis $\{\ket{\xi_{ij}}\}_{i, j=0}^1$ into the computational basis $\{\ket k\}_{k=0}^3$ of a four-dimensional system $\bO$. Finally, that computational basis is measured using a four-dimensional CNOT operation, defined by $\ket k \ket l \rightarrow \ket k \ket {l+k}$, where \enquote{+} denotes addition modulo 4. 

Now, to get the analysis of decoherence off the ground, one must begin by choosing a set $\mathfrak B$ of \textit{systems of interest}. Our chosen set $\mathfrak B_1$ will consist precisely of those systems that are labeled with black letters in \cref{eq:mmt}, i.e.\
\begin{equation}
    \mathfrak B_1 := \{ \bA, \bB, \bO, \bM_1, \bM_2\}.
\end{equation}

For an arbitrary circuit and set $\mathfrak B$ of systems of interest, \citep{ormrod2026decoherence} defines what it means for each $\bX \in \mathfrak B$ to be decohered by the combination of every system $\bY \in \mathfrak B$ in its causal future. To state the definition, we first adopt a useful representation of each system algebra $\bX$ as an algebra $\tilde \bX$ on the total input space of the circuit, obtained by pulling $\bX$ back through the unitary transformations in the circuit. In our example, we have
\begin{equation}
\begin{aligned}
\tilde \bA &:= \bA &
\qquad \qquad \qquad \qquad
\tilde \bO &:= \cu^{-1}(\bO) \\
\tilde \bB &:= \bB &
\qquad
\tilde \bM_2 &:= \cu^{-1} \circ {\rm CNOT}^{-1}(\bM_2) \\ 
\tilde \bM_1 &:= \bM_1
\end{aligned}
\end{equation}
Now, we say that $\bY$ is in the \textit{causal future} of $\bX$ if and only if $\tilde \bX$ and $\tilde \bY$ do not commute and $\bY$ comes higher up in the circuit than $\bX$. For each $\bX \in \mathfrak B$, we denote by $J_\mathfrak B^\uparrow(\bX)$ the set of all $\bY\in \mathfrak B$ in the causal future of $\bX$. The algebra representing the causal future of $\bX$ within $\mathfrak B$ is then given by
\begin{equation}
    \bJ^\uparrow_\mathfrak B(\bX) := \bigvee_{\bY \in J^\uparrow_\mathfrak B(\bX)} \tilde \bY.
\end{equation}
We then let $\bJ^\uparrow_\mathfrak B(\bX)$ play the role of the environment fragment $\bG$ from the previous subsection: generalizing \cref{eq:dec}, the \textit{decohered subalgebra} $\tilde \bX^\uparrow_{{\rm dec}\mathfrak B}$ of $\tilde \bX$ is defined by
\begin{equation} \label{eq:gen_dec}
    \tilde \bX^\uparrow_{{\rm dec}\mathfrak B}
    := \tilde \bX^\uparrow_{{\rm sur}\mathfrak B}
    \cap \tilde \bX^\uparrow_{{\rm rep}\mathfrak B},
\end{equation}
where
\begin{equation} \label{eq:gen_sur_rep}
\tilde \bX^\uparrow_{{\rm sur}\mathfrak B}
:= \bJ^\uparrow_\mathfrak B(\bX)' \cap  \tilde \bX
\qquad\text{and}\qquad
\tilde \bX^\uparrow_{{\rm rep}\mathfrak B}
:= \bigl(\bJ^\uparrow_\mathfrak B(\bX)' \cap \tilde \bX\bigr)'
\cap \tilde \bX .
\end{equation}
$\tilde \bX^\uparrow_{{\rm dec} \mathfrak B}$ can then be pushed forward through the circuit to obtain a representation of the decohered subalgebra as a subalgebra $\bX^\uparrow_{{\rm dec} \mathfrak B}$ of $\bX$.

As dedicated readers may wish to verify for themselves, in the PBR example of \cref{eq:mmt} we obtain the following five decohered algebras,
\begin{equation}
\begin{alignedat}{2}
\bA^\uparrow_{{\rm dec}\mathfrak B_1}
&= \mathbb C I,
\qquad\qquad \qquad\qquad
&
\bB^\uparrow_{{\rm dec}\mathfrak B_1}
&= \mathbb C I,
\\
\bO^\uparrow_{{\rm dec}\mathfrak B_1}
&= \gen{Z_\bO},
\qquad\qquad
&
\bM^\uparrow_{1\,{\rm dec}\mathfrak B_1}
&= \mathbb C I,
\\
\bM^\uparrow_{2\,{\rm dec}\mathfrak B_1}
&= \mathbb C I,
\end{alignedat}
\end{equation}
where $Z_\bO$ is the four-dimensional Pauli operator $Z_\bO=\sum_{k=0}^3 i^k \ket k \bra k$ and the algebra it generates is $\gen{Z_\bO} = {\rm span}_\mathbb C\{\ket k \bra k_\bO\}_{k=0}^3$. This derived $\bO^\uparrow_{{\rm dec}\mathfrak B_1}$ is consistent with our intuition that in this model $\bA\bB$ are measured in the PBR basis $\{\ket{\xi_{ij}}\}_{i, j=0}^1$ (which we recall is transformed to $\{\ket{k}\}_{k=0}^3$ by $\cu$).
While in this example there is only one nontrivial decohered algebra, we will soon see that in other examples there can be many.

An important feature of this theory of decoherence is that different decohered subalgebras are compatible with each other, and can thus be consistently \enquote{glued together}. For any unitary circuit and set $\mathfrak B$ of systems, all decohered algebras combine to form a single \textit{commutative} algebra\footnote{The proof that $\mathfrak B^\uparrow_{\rm dec}$ is commutative is quite simple: for any pair of systems $\bX, \bY \in \mathfrak B$, if $\tilde \bX$ and $\tilde \bY$ do not commute with each other, then one of the two systems must be in the causal future of the other. Assuming without loss of generality that $\bY$ is in the causal future of $\bX$, by definition $\tilde \bX^\uparrow_{{\rm dec}\mathfrak B}$ commutes with $\tilde \bY$, and thus $[\tilde \bX^\uparrow_{{\rm dec} \mathfrak B}, \tilde \bY^\uparrow_{{\rm dec} \mathfrak B}] = 0$. Since all of the decohered algebras commute with each other, as well as being commutative in their own right, it follows that the algebra they generate is commutative too.}
\begin{equation} \label{eq:big_dec}
    \mathfrak B^\uparrow_{\rm dec}
    := \bigvee_{\bX \in \mathfrak B}
    \tilde \bX^\uparrow_{{\rm dec}\mathfrak B}.
\end{equation}
Being commutative, $\mathfrak B^\uparrow_{\rm dec}$ is spanned by a preferred PVM $\tilde \pi^{H^\uparrow}$ on the total Hilbert space. This \textit{history PVM} can be written as a product of each individual system PVM,
\begin{equation}
    \tilde \pi^{H^\uparrow} = \prod_{\bX \in \mathfrak B}\tilde \pi^{X^\uparrow}_\bX,
\end{equation}
where the \textit{history variable} $H^\uparrow$ is defined as $H^\uparrow:= \bigtimes_{\bX \in \mathfrak B} X^\uparrow$. (We note that $\tilde \pi^{h^\uparrow}$ may be zero for some $h^\uparrow \in H^\uparrow$.)

So far, we have been considering only how systems get decohered by other systems in their future. But since the unitary circuit is reversible, the distinction between \enquote{past} and \enquote{future} is a matter of convention. We can therefore carry out the same analysis of decoherence with respect to each system's causal past, defining $\bJ^\downarrow(\bX)$ as the algebra generated by the set containing $\tilde \bY$ for each $\bY \in \mathfrak B$ in the causal past of $\bX$, and defining the \textit{dual decohered} algebra $\tilde \bX^\downarrow_{{\rm dec}\mathfrak B}$ by expressions that exactly resemble \cref{eq:gen_dec} and \cref{eq:gen_sur_rep} except that each $\uparrow$ is replaced with $\downarrow$. Just like the decohered case, all dual decohered algebras glue together to form a single commutative algebra
\begin{equation} \label{eq:big_dual}
    \mathfrak B^\downarrow_{\rm dec} := \bigvee_{\bX \in \mathfrak B} \tilde \bX^\downarrow_{{\rm dec} \mathfrak B}
\end{equation}
spanned by a history PVM
\begin{equation}
    \tilde \epsilon^{H^\downarrow} = \prod_{\bX \in \mathfrak B}\tilde \epsilon_\bX^{X^\downarrow}.
\end{equation}
We will see some examples of dual decohered algebras in the next subsection.

That wraps up our discussion of the theory of the dynamics of decoherence underpinning DFQT.
For what is to come, the key thing to remember is that it was the unitary dynamics alone that marked out the two history PVMs $\tilde \pi^{H^\uparrow}$ and $\tilde \epsilon^{H^\downarrow}$ as special---our analysis assumed nothing about the quantum state fed into the bottom of the circuit.
Indeed, as we are about to see, we do not have to believe that there is such a state at all.

\subsection{How to put the dynamics first} \label{subsec:first}

Some state-first assumptions are so common that they are often not consciously recognized as assumptions: like air, they go easily unnoticed.
Here, the state-first assumption that needs to be called out and resisted is that the unitary circuit acts on some initial state fed into the bottom of it. 
That is how circuits are usually interpreted, but that is not their role in DFQT. 

In DFQT, the unitary circuit enjoys primitive ontological status,\footnote{One might object that a unitary transformation cannot be thought of as ontologically primitive because mathematically it is a function. But a density operator is a function on Hilbert-space vectors, and this is not usually taken as a reason that it cannot represent a state. Taking the unitary dynamics as primitive does not require us to stop representing them as a function---it just requires us to think differently about what that function means.} and all physical states emerge from it.
These physical states are constituted by events, and the mechanism through which they emerge is decoherence and dual decoherence. Quantum states are not physical states, but useful tools for predicting the probability of one collection of events conditioned on another. 

Events in the conditioning set associated with some quantum state will often include some emerging from an interaction that we might ordinarily describe as preparing the system in that quantum state. Accordingly, we will now extend the circuit model of the PBR scenario from  \cref{eq:mmt} to include the physical interactions by which $\bA$ and $\bB$ are prepared. 
As in \cref{sec:pbr}, we denote by $\ket{\psi_0}$ and $\ket{\psi_1}$ the states $\ket 0$ and $\frac{1}{\sqrt{2}}(\ket 0 + \ket 1)$ respectively, where $\{\ket 0, \ket 1\}$ is an orthonormal basis.
We will assume that for each of $\bA$ and $\bB$, the state $\ket{\psi_i}$ is prepared in two stages: first $\ket{\psi_0}$ is prepared, and then it is either left alone or transformed into $\ket{\psi_1}$. 
The first stage will be represented by a CNOT. The second will be represented by the controlled-Hadamard transformation, defined by the unitary operator $\ket 0\bra 0 \otimes I + \ket 1 \bra 1 \otimes H$ where $H \ket {\psi_i} = \ket {\psi_{i+1}}$ and \enquote{+} denotes addition modulo two. This leaves us with the following circuit:
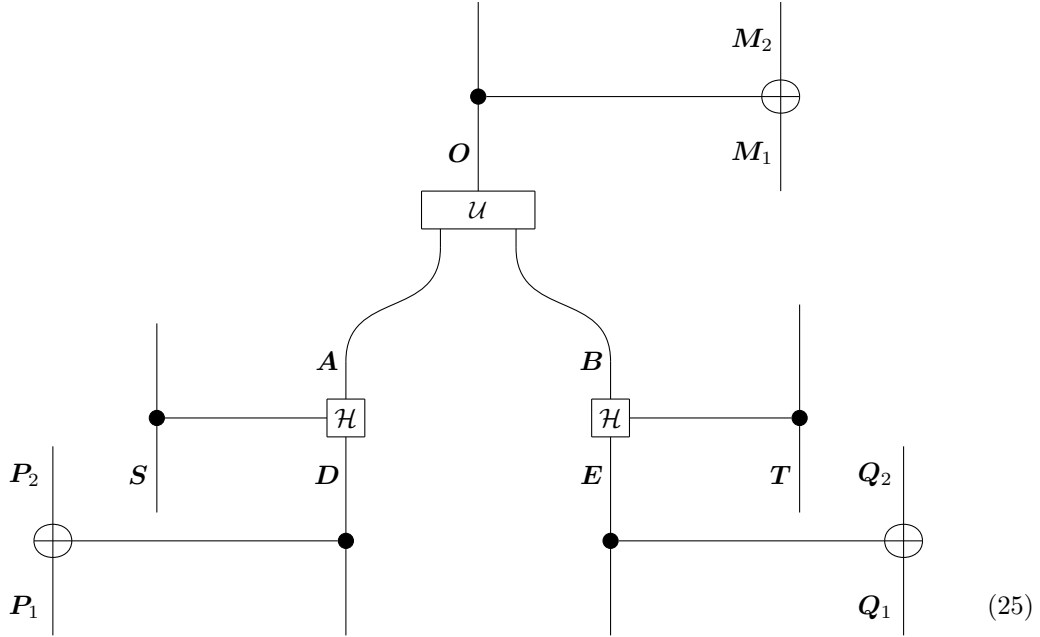
\begin{equation} \label{eq:pbr_circ}
    \begin{tikzpicture}
	\begin{pgfonlayer}{nodelayer}
		\node [style=none] (0) at (9, -0.75) {};
		\node [style=none] (1) at (9, 4.5) {};
		\node [style=none] (2) at (1.25, -0.75) {};
		\node [style=none] (3) at (8.5, 4.5) {};
		\node [style=none] (4) at (9.5, 4.5) {};
		\node [style=none] (5) at (9.5, 5.5) {};
		\node [style=none] (6) at (8.5, 5.5) {};
		\node [style=none] (7) at (9, 5) {$\ch$};
		\node [style=none] (8) at (8.5, 5) {};
		\node [style=black dot] (9) at (4, 5) {};
		\node [style=none] (10) at (4, 2.5) {};
		\node [style=none] (11) at (4, 7.5) {};
		\node [style=none] (12) at (9, 5.5) {};
		\node [style=none] (13) at (1.25, 4.25) {};
		\node [style=none] (14) at (0.5, 0) {$\bP_1$};
		\node [style=none] (15) at (8.5, 3.5) {$\bD$};
		\node [style=none] (16) at (9, 6.5) {};
		\node [style=none] (17) at (11.5, 9.5) {};
		\node [style=none] (18) at (20.5, 11) {};
		\node [style=none] (19) at (20.5, 16) {};
		\node [style=none] (20) at (11, 10) {};
		\node [style=none] (21) at (14, 10) {};
		\node [style=none] (22) at (14, 11) {};
		\node [style=none] (23) at (11, 11) {};
		\node [style=none] (24) at (12.5, 11) {};
		\node [style=none] (25) at (12.5, 16) {};
		\node [style=black dot] (26) at (12.5, 13.5) {};
		\node [style=none] (27) at (12.5, 10.5) {\small $\cu$};
		\node [style=none] (28) at (12, 12) {$\bO$};
		\node [style=none] (29) at (8.5, 6.5) {$\bA$};
		\node [style=none] (30) at (19.75, 15) {$\bM_2$};
		\node [style=none] (31) at (11.5, 10) {};
		\node [style=none] (32) at (9, 5.5) {};
		\node [style=none] (33) at (16, 6.5) {};
		\node [style=none] (34) at (13.5, 9.5) {};
		\node [style=none] (35) at (15.5, 6.5) {$\bB$};
		\node [style=none] (36) at (13.5, 10) {};
		\node [style=none] (37) at (16, 5.5) {};
		\node [style=none] (38) at (21, 13.5) {};
		\node [style=none] (39) at (20, 13.5) {};
		\node [style=none] (40) at (21, 13.5) {};
		\node [style=none] (41) at (20, 13.5) {};
		\node [style=black dot] (42) at (9, 1.75) {};
		\node [style=none] (43) at (0.75, 1.75) {};
		\node [style=none] (44) at (1.75, 1.75) {};
		\node [style=none] (45) at (0.75, 1.75) {};
		\node [style=none] (46) at (1.75, 1.75) {};
		\node [style=none] (47) at (0.5, 3.5) {$\bP_2$};
		\node [style=none] (48) at (16, -0.75) {};
		\node [style=none] (49) at (16, 4.5) {};
		\node [style=none] (50) at (23.75, -0.75) {};
		\node [style=none] (51) at (16.5, 4.5) {};
		\node [style=none] (52) at (15.5, 4.5) {};
		\node [style=none] (53) at (15.5, 5.5) {};
		\node [style=none] (54) at (16.5, 5.5) {};
		\node [style=none] (55) at (16, 5) {$\ch$};
		\node [style=none] (56) at (16.5, 5) {};
		\node [style=black dot] (57) at (21, 5) {};
		\node [style=none] (58) at (21, 2.5) {};
		\node [style=none] (59) at (21, 8) {};
		\node [style=none] (60) at (16, 5.5) {};
		\node [style=none] (61) at (23.75, 4.25) {};
		\node [style=none] (62) at (20.5, 3.5) {$\bT$};
		\node [style=none] (63) at (23, 0) {$\bQ_1$};
		\node [style=none] (64) at (15.5, 3.5) {$\bE$};
		\node [style=none] (65) at (16, 5.5) {};
		\node [style=black dot] (66) at (16, 1.75) {};
		\node [style=none] (67) at (24.25, 1.75) {};
		\node [style=none] (68) at (23.25, 1.75) {};
		\node [style=none] (69) at (24.25, 1.75) {};
		\node [style=none] (70) at (23.25, 1.75) {};
		\node [style=none] (71) at (23, 3.5) {$\bQ_2$};
		\node [style=none] (72) at (3.5, 3.5) {$\bS$};
		\node [style=none] (73) at (19.75, 12) {$\bM_1$};
	\end{pgfonlayer}
	\begin{pgfonlayer}{edgelayer}
		\draw (1.center) to (0.center);
		\draw (6.center) to (3.center);
		\draw (3.center) to (4.center);
		\draw (4.center) to (5.center);
		\draw (5.center) to (6.center);
		\draw (9) to (8.center);
		\draw (11.center) to (10.center);
		\draw [in=90, out=-90, looseness=1.25] (17.center) to (16.center);
		\draw (23.center) to (20.center);
		\draw (20.center) to (21.center);
		\draw (21.center) to (22.center);
		\draw (22.center) to (23.center);
		\draw (24.center) to (25.center);
		\draw (31.center) to (17.center);
		\draw (32.center) to (16.center);
		\draw [in=90, out=-90, looseness=1.25] (34.center) to (33.center);
		\draw (36.center) to (34.center);
		\draw (37.center) to (33.center);
		\draw (26) to (38.center);
		\draw (19.center) to (18.center);
		\draw [bend left=90, looseness=1.50] (39.center) to (38.center);
		\draw [bend right=90, looseness=1.50] (41.center) to (40.center);
		\draw (42) to (43.center);
		\draw [bend right=90, looseness=1.50] (44.center) to (43.center);
		\draw [bend left=90, looseness=1.50] (46.center) to (45.center);
		\draw (2.center) to (13.center);
		\draw (49.center) to (48.center);
		\draw (54.center) to (51.center);
		\draw (51.center) to (52.center);
		\draw (52.center) to (53.center);
		\draw (53.center) to (54.center);
		\draw (57) to (56.center);
		\draw (59.center) to (58.center);
		\draw (66) to (67.center);
		\draw [bend left=90, looseness=1.50] (68.center) to (67.center);
		\draw [bend right=90, looseness=1.50] (70.center) to (69.center);
		\draw (50.center) to (61.center);
	\end{pgfonlayer}
\end{tikzpicture}
\end{equation}

Recall that the analysis of decoherence begins with choosing a set of systems. Our chosen set $\mathfrak B_2$ will be the set of all labelled systems, i.e.\ 
\begin{equation} \label{eq:bubble2}
 \mathfrak B_2 = \{\bP_1, \bP_2, \bQ_1, \bQ_2, \bS, \bD, \bE, \bT, \bA, \bB, \bO, \bM_1, \bM_2\}.
\end{equation}
This is a set of thirteen systems, so technically there will be twenty-six algebras that decohere or dual decohere relative to it. However, most of these are trivial. In fact, there are only five nontrivial decohered or dual decohered algebras,
\begin{equation} \label{eq:pbr_algs}
\begin{alignedat}{2}
\bS^\uparrow_{{\rm dec}\mathfrak B_2}
&= \gen{Z_\bS}
\qquad\qquad \qquad\qquad
&
\bD^\downarrow_{{\rm dec}\mathfrak B_2}
&= \gen{Z_\bD}
\\
\bT^\uparrow_{{\rm dec}\mathfrak B_2}
&= \gen{Z_\bT}
\qquad\qquad \qquad\qquad
&
\bE^\downarrow_{{\rm dec}\mathfrak B_2}
&= \gen{Z_\bE}
\\
\bO^\uparrow_{{\rm dec}\mathfrak B_2}
&= \gen{Z_\bO},
\end{alignedat}
\end{equation}
where each $Z_\bX$ represents a Pauli operator (which in the case of $\bO$ is the four-dimensional generalized Pauli $Z_\bO=\sum_{k=0}^3 i^k \ket k \bra k$). For visualization, we depict these preferred algebras on the circuit from which they emerged,
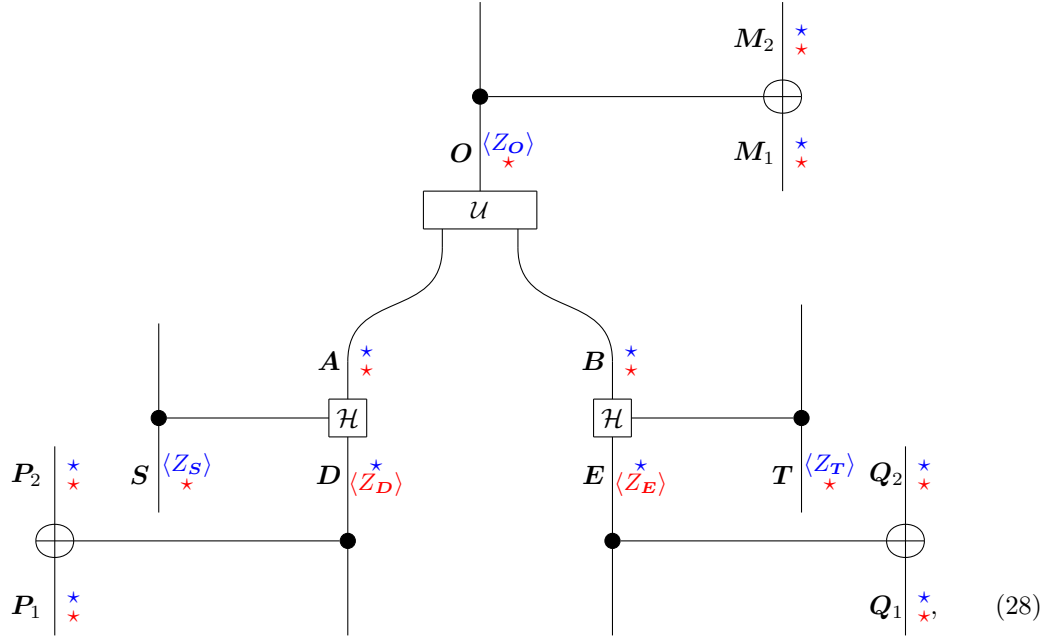
\begin{equation} \label{eq:pbr_circ_decorated}
    \begin{tikzpicture}
	\begin{pgfonlayer}{nodelayer}
		\node [style=none] (0) at (9, -0.75) {};
		\node [style=none] (1) at (9, 4.5) {};
		\node [style=none] (2) at (1.25, -0.75) {};
		\node [style=none] (3) at (8.5, 4.5) {};
		\node [style=none] (4) at (9.5, 4.5) {};
		\node [style=none] (5) at (9.5, 5.5) {};
		\node [style=none] (6) at (8.5, 5.5) {};
		\node [style=none] (7) at (9, 5) {$\ch$};
		\node [style=none] (8) at (8.5, 5) {};
		\node [style=black dot] (9) at (4, 5) {};
		\node [style=none] (10) at (4, 2.5) {};
		\node [style=none] (11) at (4, 7.5) {};
		\node [style=none] (12) at (9, 5.5) {};
		\node [style=none] (13) at (1.25, 4.25) {};
		\node [style=none] (14) at (0.5, 0) {$\bP_1$};
		\node [style=none] (15) at (8.5, 3.5) {$\bD$};
		\node [style=none] (16) at (9.75, 3.25) {\color{red}\small $\gen{Z_\bD}$};
		\node [style=none] (17) at (9.75, 3.75) {\color{blue}\small $\star$};
		\node [style=none] (18) at (1.75, 3.75) {\color{blue} \small $\star$};
		\node [style=none] (19) at (1.75, 0.25) {\color{blue}\small $\star$};
		\node [style=none] (20) at (4.75, 3.75) {\color{blue}\small $\gen{Z_\bS}$};
		\node [style=none] (21) at (4.75, 3.25) {\color{red}\small $\star$};
		\node [style=none] (22) at (9, 6.5) {};
		\node [style=none] (23) at (11.5, 9.5) {};
		\node [style=none] (24) at (20.5, 11) {};
		\node [style=none] (25) at (20.5, 16) {};
		\node [style=none] (26) at (11, 10) {};
		\node [style=none] (27) at (14, 10) {};
		\node [style=none] (28) at (14, 11) {};
		\node [style=none] (29) at (11, 11) {};
		\node [style=none] (30) at (12.5, 11) {};
		\node [style=none] (31) at (12.5, 16) {};
		\node [style=black dot] (32) at (12.5, 13.5) {};
		\node [style=none] (33) at (12.5, 10.5) {\small $\cu$};
		\node [style=none] (34) at (12, 12) {$\bO$};
		\node [style=none] (35) at (8.5, 6.5) {$\bA$};
		\node [style=none] (36) at (19.75, 15) {$\bM_2$};
		\node [style=none] (37) at (11.5, 10) {};
		\node [style=none] (38) at (9, 5.5) {};
		\node [style=none] (39) at (16, 6.5) {};
		\node [style=none] (40) at (13.5, 9.5) {};
		\node [style=none] (41) at (15.5, 6.5) {$\bB$};
		\node [style=none] (42) at (13.5, 10) {};
		\node [style=none] (43) at (16, 5.5) {};
		\node [style=none] (44) at (21, 13.5) {};
		\node [style=none] (45) at (20, 13.5) {};
		\node [style=none] (46) at (21, 13.5) {};
		\node [style=none] (47) at (20, 13.5) {};
		\node [style=none] (48) at (9.5, 6.75) {\color{blue}\small $\star$};
		\node [style=none] (49) at (16.5, 6.75) {\color{blue}\small $\star$};
		\node [style=none] (50) at (13.25, 12.25) {\color{blue}\small $\gen{Z_\bO}$};
		\node [style=none] (51) at (21, 15.25) {\color{blue}\small $\star$};
		\node [style=black dot] (52) at (9, 1.75) {};
		\node [style=none] (53) at (0.75, 1.75) {};
		\node [style=none] (54) at (1.75, 1.75) {};
		\node [style=none] (55) at (0.75, 1.75) {};
		\node [style=none] (56) at (1.75, 1.75) {};
		\node [style=none] (57) at (0.5, 3.5) {$\bP_2$};
		\node [style=none] (58) at (16, -0.75) {};
		\node [style=none] (59) at (16, 4.5) {};
		\node [style=none] (60) at (23.75, -0.75) {};
		\node [style=none] (61) at (16.5, 4.5) {};
		\node [style=none] (62) at (15.5, 4.5) {};
		\node [style=none] (63) at (15.5, 5.5) {};
		\node [style=none] (64) at (16.5, 5.5) {};
		\node [style=none] (65) at (16, 5) {$\ch$};
		\node [style=none] (66) at (16.5, 5) {};
		\node [style=black dot] (67) at (21, 5) {};
		\node [style=none] (68) at (21, 2.5) {};
		\node [style=none] (69) at (21, 8) {};
		\node [style=none] (70) at (16, 5.5) {};
		\node [style=none] (71) at (23.75, 4.25) {};
		\node [style=none] (72) at (20.5, 3.5) {$\bT$};
		\node [style=none] (73) at (23.25, 0) {$\bQ_1$};
		\node [style=none] (74) at (15.5, 3.5) {$\bE$};
		\node [style=none] (75) at (16.75, 3.25) {\color{red} \small $\gen{Z_\bE}$};
		\node [style=none] (76) at (16.75, 3.75) {\color{blue}\small $\star$};
		\node [style=none] (77) at (24.25, 3.75) {\color{blue} \small $\star$};
		\node [style=none] (78) at (24.25, 0.25) {\color{blue}\small $\star$};
		\node [style=none] (79) at (21.75, 3.75) {\color{blue}\small $\gen{Z_\bT}$};
		\node [style=none] (80) at (21.75, 3.25) {\color{red}\small $\star$};
		\node [style=none] (81) at (16, 5.5) {};
		\node [style=black dot] (82) at (16, 1.75) {};
		\node [style=none] (83) at (24.25, 1.75) {};
		\node [style=none] (84) at (23.25, 1.75) {};
		\node [style=none] (85) at (24.25, 1.75) {};
		\node [style=none] (86) at (23.25, 1.75) {};
		\node [style=none] (87) at (23.25, 3.5) {$\bQ_2$};
		\node [style=none] (88) at (24.25, -0.25) {\color{red}\small $\star$};
		\node [style=none] (89) at (1.75, -0.25) {\color{red}\small $\star$};
		\node [style=none] (90) at (13.25, 11.75) {\color{red}\small $\star$};
		\node [style=none] (91) at (21, 14.75) {\color{red}\small $\star$};
		\node [style=none] (92) at (16.5, 6.25) {\color{red}\small $\star$};
		\node [style=none] (93) at (9.5, 6.25) {\color{red}\small $\star$};
		\node [style=none] (94) at (3.5, 3.5) {$\bS$};
		\node [style=none] (95) at (24.25, 3.25) {\color{red}\small $\star$};
		\node [style=none] (96) at (1.75, 3.25) {\color{red}\small $\star$};
		\node [style=none] (97) at (19.75, 12) {$\bM_1$};
		\node [style=none] (98) at (21, 12.25) {\color{blue}\small $\star$};
		\node [style=none] (99) at (21, 11.75) {\color{red}\small $\star$};
	\end{pgfonlayer}
	\begin{pgfonlayer}{edgelayer}
		\draw (1.center) to (0.center);
		\draw (6.center) to (3.center);
		\draw (3.center) to (4.center);
		\draw (4.center) to (5.center);
		\draw (5.center) to (6.center);
		\draw (9) to (8.center);
		\draw (11.center) to (10.center);
		\draw [in=90, out=-90, looseness=1.25] (23.center) to (22.center);
		\draw (29.center) to (26.center);
		\draw (26.center) to (27.center);
		\draw (27.center) to (28.center);
		\draw (28.center) to (29.center);
		\draw (30.center) to (31.center);
		\draw (37.center) to (23.center);
		\draw (38.center) to (22.center);
		\draw [in=90, out=-90, looseness=1.25] (40.center) to (39.center);
		\draw (42.center) to (40.center);
		\draw (43.center) to (39.center);
		\draw (32) to (44.center);
		\draw (25.center) to (24.center);
		\draw [bend left=90, looseness=1.50] (45.center) to (44.center);
		\draw [bend right=90, looseness=1.50] (47.center) to (46.center);
		\draw (52) to (53.center);
		\draw [bend right=90, looseness=1.50] (54.center) to (53.center);
		\draw [bend left=90, looseness=1.50] (56.center) to (55.center);
		\draw (2.center) to (13.center);
		\draw (59.center) to (58.center);
		\draw (64.center) to (61.center);
		\draw (61.center) to (62.center);
		\draw (62.center) to (63.center);
		\draw (63.center) to (64.center);
		\draw (67) to (66.center);
		\draw (69.center) to (68.center);
		\draw (82) to (83.center);
		\draw [bend left=90, looseness=1.50] (84.center) to (83.center);
		\draw [bend right=90, looseness=1.50] (86.center) to (85.center);
		\draw (60.center) to (71.center);
	\end{pgfonlayer}
\end{tikzpicture},
\end{equation}
where $\star:= \mathbb C I$ and where we are colour-coding {\color{blue} decohered} and {\color{red} dual decohered} algebras for additional visual clarity.

What we now need to do is make precise the sense in which these algebras define emergent events. An \textit{event} is the selection of a unique subspace in the preferred subspace decomposition induced by some individual decohered or dual decohered algebra; formally, it can be thought of as a variable {\color{blue} $X^\uparrow$} or {\color{red} $X^\downarrow$} taking a value. Because nothing in the circuit predetermines which value this will be, DFQT is an indeterministic theory, and therefore requires a probability distribution specifying the likelihood of a given collection of events to happen. 

There is a very natural candidate for what this probability distribution should be. Recall that for any circuit and set $\mathfrak B$ of systems of interest, the families of decohered and dual decohered PVMs each combine to form a single history PVM, $\tilde \pi^{{\color{blue} H^\uparrow}}= \prod_{\bX \in \mathfrak B}\tilde \pi_\bX^{{\color{blue} X^\uparrow}}$ and $\tilde \epsilon^{{\color{red} H^\downarrow}}= \prod_{\bX \in \mathfrak B}\tilde \epsilon_\bX^{{\color{red} X^\downarrow}}$ respectively. Like any pair of PVMs, the two history PVMs define a probability distribution by tracing their product and dividing by the dimension $d$ of the total Hilbert space, i.e.\ ${\rm Prob}({\color{blue}H^\uparrow}, {\color{red} H^\downarrow}) =  \frac{1}{d} {\rm Tr} (\tilde \epsilon^{\color{red}H^\downarrow} \tilde \pi^{\color{blue}H^\uparrow})$, where $d$ is the dimension of the total Hilbert space. Writing this probability distribution explicitly in terms of the original event-variables gives us DFQT's probability rule,
\begin{equation} \label{eq:prob_general}
    {\rm Prob}_\mathfrak B({\color{red} X_1^\downarrow, \ldots, X_N^\downarrow}{{\color{blue}, X_1^\uparrow, \ldots, X_N^\uparrow}}) = \frac{1}{d} {\rm Tr}( \tilde \epsilon^{{{\color{red} X_1^\downarrow \ldots X_N^\downarrow}}} \tilde \pi^{{{\color{blue} X_1^\uparrow \ldots X_N^\uparrow}}}),
\end{equation}
where $N$ is the number of systems in $\mathfrak B$.\footnote{As explained in \citep{ormrod2026decoherence}, this means that for any unitary circuit and any set of systems of interest, decoherence and dual decoherence select a unique privileged consistent history set \citep{griffiths1984consistent, gell1990complexity, griffiths2003consistent}, a result that addresses the history set-selection problem noted in \citep{dowker1996consistent}.}

Returning to the PBR example, evaluating \cref{eq:prob_general} for the set $\mathfrak B_2$ of systems of interest shows that
\begin{equation} \label{eq:pbr_conditional}
     {\rm Prob}_{\mathfrak B_2} ({\color{blue} O^\uparrow} | {\color{red} D^\downarrow} = {\color{red} E^\downarrow} ={\color{red}0},  {\color{blue} S^\uparrow, T^\uparrow})  = |\braket{\xi_{{\color{blue} O^\uparrow}} | \psi_{{\color{blue} S^\uparrow}},\psi_{{\color{blue} T^\uparrow}}}|^2,
\end{equation}
where the conditional distribution is defined as usual via Bayes' rule. Here, each value of ${\color{blue}O^\uparrow}$ corresponds to a unique projector on $\bO$, which is related by $\cu$ to some unique $\ket{\xi_{ij}}\bra{\xi_{ij}}_{\bA\bB}$; we are labelling the value of ${\color{blue}O^\uparrow}$ with the corresponding indices $i, j$ to match the labels for the PBR measurement outcomes from \cref{sec:pbr}.

\cref{eq:pbr_conditional} shows that whenever the events ${\color{red} D^\downarrow} = {\color{red} 0}$ and ${\color{red} E^\downarrow} = {\color{red} 0}$ take place, the standard quantum predictions for the PBR scenario are recovered exactly. 
This justifies us saying that, when ${\color{red} D^\downarrow} = {\color{red} 0}$ and ${\color{red} E^\downarrow} = {\color{red} 0}$, the systems $\bA$ and $\bB$ are successfully prepared in the states $\ket{\psi_{\color{blue}S^\uparrow}}$ and $\ket{\psi_{\color{blue}T^\uparrow}}$. 
However, this statement should not be interpreted as saying that $\bA$ and $\bB$ are prepared in physical states represented by $\ket{\psi_{\color{blue}S^\uparrow}}$ and $\ket{\psi_{\color{blue}T^\uparrow}}$, nor even that they are prepared in some physical state on which their quantum states supervene. 
All we mean is that $\bA$ and $\bB$ have been prepared in such a way that assigning $\ket{\psi_{\color{blue}S^\uparrow}}$ and $\ket{\psi_{\color{blue}T^\uparrow}}$ allows us to calculate the correct probabilities. 

Having laid out its formalism and seen an example, we can now give a final summary of DFQT. 
This is a theory in which ontologically primitive unitary dynamics give birth to events via decoherence and dual decoherence.
The events emerge stochastically, following the probability rule in \cref{eq:prob_general}.
And they do so relationally: because decoherence and dual decoherence are defined relative to a subset $\mathfrak B$ of systems in the circuit, events take place only relative to such subsets. In DFQT as currently formulated, events emerging relative to different sets of systems are probabilistically independent, though the possibility of a successor theory allowing cross-set correlations is a topic of ongoing research.

Recall that in state-first theories, dynamics are reducible to counterfactual relations of dependence between states, whereas in dynamics-first theories the dynamics do not supervene on such relations. In DFQT, the supervenience claim does indeed fail: there exist distinct unitary circuits that give rise to exactly the same emergent events and probabilities. For example, the circuits consisting respectively of just the identity $\ci: \bA \rightarrow \bB$ from qubit $\bA$ to qubit $\bB$ and just the Pauli unitary channel $\cx: \bA \rightarrow \bB$ both give rise to no nontrivial events at all, and thus to a trivial probability distribution.\footnote{\label{ft:leibniz} Note, however, that this does not imply that there are features of the dynamical ontology that have no operational consequence; to the contrary, any two distinct dynamical transformations can be operationally distinguished by embedding them in a larger circuit in which the surrounding dynamics are modified. For example, suppose we enlarge the circuits consisting of $\ci$ and $\cx$ in each the same way, by adding two CNOTs, one controlled on $\bA$ and the other on $\bB$. In both of the enlarged circuits, there are $Z$-events emerging at $\bA$ and $\bB$ (relative to a certain set of systems), but these events are correlated in the first circuit and anticorrelated in the second.}

Evaluating DFQT is not the main purpose of this work, but it is worth giving one reason why it is worth taking seriously. Everett's many worlds interpretation is appealing to many because it promises a way of understanding bare unitary quantum theory as an observer-independent description of reality. One of its biggest challenges, however, is making sense of the role of probability in quantum theory with its ontology of an unsupplemented and deterministically evolving quantum state. Like the Everett interpretation, DFQT promises a precise, realist interpretation of unitary quantum theory. However, it does not appear to face any particular problem with probabilities. Indeed, since DFQT's probabilities represent the chance of genuinely stochastic events, it has no more of a problem with probabilities than objective collapse theories.

\section{Reconciliation} \label{sec:reconciliation}

The notion of $\psi$-ontology formalized by \citep{harrigan2010einstein} and commonly thought to be established by PBR is that the pure quantum state assigned to a system supervenes on that system's physical state. But our DFQT model of the PBR scenario denies this.

To see that it does, first note that the quantum state $\ket{\psi_{\color{blue}s^\uparrow}}_\bA$, which is assigned to $\bA$ to help compute conditional probabilities of events emerging relative to $\mathfrak B_2$, does not supervene on the physical events that emerge at $\bA$ relative to $\mathfrak B_2$. Indeed, it cannot, because these events are trivial: $\bA^\uparrow_{{\rm dec}\mathfrak B_2}$ and $\bA^\downarrow_{{\rm dec}\mathfrak B_2}$ are each equal to $\mathbb C I$, and thus the associated event-variables ${\color{blue} A^\uparrow}$ and ${\color{red} A^\downarrow}$ can each take only one possible value. Supervenience must then fail because any function $f: {\color{blue} A^\uparrow} \times {\color{red} A^\downarrow} \rightarrow \{\ket{\psi_0}_\bA, \ket{\psi_1}_\bA\}$ can only assign a fixed quantum state, whereas in the experiment we are considering $\ket{\psi_0}_\bA$ is prepared on some runs and $\ket{\psi_1}_\bA$ on others. 

Nor does the assigned quantum state supervene on events at $\bA$ that emerge relative to other sets of systems. Indeed, the quantum state $\ket{\psi_{\color{blue}s^\uparrow}}_\bA$ is usefully assigned to $\bA$ precisely when the events ${\color{red} D^\downarrow}={\color{red} 0}$ and ${\color{blue} S^\uparrow}={\color{blue}s^\uparrow}$ emerge relative to $\mathfrak B_2$. It follows that supervenience would require correlations between these events and events at $\bA$ emerging relative to other sets $\mathfrak B$ on which the quantum state is supposed to supervene. But according to DFQT, there are no such cross-set correlations.\footnote{As noted in \cref{sec:cdt}, the possibility of a successor theory to DFQT that does posit cross-set correlations is an open question in the dynamics-first research programme. But even in such a theory, it seems unlikely that the supervenience claim will hold: in our PBR model, the only nontrivial subalgebra of $\bA$ that decoheres or dual decoheres relative to any $\mathfrak B$ is the one generated by $(\ket{\psi_0}_\bA+\ket{\psi_1}_\bA)(\bra{\psi_0}_\bA+\bra{\psi_1}_\bA)$. Since this algebra defines an event space that is incompatible with those generated by each $\ket{\psi_{\color{blue}s^\uparrow}}$, one does not naturally expect the perfect cross-set correlations required for supervenience.} 

In DFQT, the physical state $\Lambda_\bA^\mathfrak B$ of $\bA$ \textit{relative} to $\mathfrak B$ is naturally defined as the pair of events that emerge at $\bA$ relative to $\mathfrak B$, from decoherence and dual decoherence respectively. Then the \textit{absolute} physical state $\Lambda_\bA$ of $\bA$ can be defined as the collection of all relative states $\Lambda_\bA:= \{\Lambda_\bA^\mathfrak B\}_\mathfrak B$. What the preceding arguments show is that the assigned quantum state $\ket{\psi_{\color{blue}s^\uparrow}}_\bA$ supervenes on neither $\Lambda_\bA^{\mathfrak B_2}$ nor $\Lambda_\bA$. Hence, according to DFQT, $\psi$-ontology is false.

It is worth clarifying that the quantum state assigned to $\bA$ does supervene on events that emerge at \textit{other} systems, namely ${\color{blue} S^\uparrow}$ and ${\color{red} D^\downarrow}$. However, this fact does not establish $\psi$-ontology. Indeed, many (if not most) antirealists about the quantum state will readily admit that the assigned quantum state supervenes on \textit{some} objective physical facts, even if only the setting on the device used to prepare the system. For precisely this reason, \citep{harrigan2010einstein} was careful to define $\psi$-ontology as the specific claim that the quantum state supervenes on physical facts \textit{about the system to which it is assigned}.

The way that DFQT reconciles the denial of $\psi$-ontology with the PBR correlations is by denying that the physical state of a system screens off the preparation settings from the measurement outcome. 
Indeed, the physical state $\Lambda_{\bA\bB}^{\mathfrak B_2}$ of $\bA\bB$ relative to $\mathfrak B_2$, defined as the list of all events emerging at $\bA$ and $\bB$ relative to $\mathfrak B_2$, is trivial, and therefore cannot screen off any nontrivial correlations. On the other hand, PBR's ontological separability and preparation independence assumptions are trivially satisfied by $\Lambda_{\bA\bB}^{\mathfrak B_2}$. By the absence of cross-set correlations, the absolute physical state $\Lambda_{\bA\bB}$ also violates the screening-off condition. 

From a state-first point of view, it is tempting to make the following assessment of our PBR model: the settings ${\color{blue}S^\uparrow}$ and ${\color{blue}T^\uparrow}$ must surely influence the outcome ${\color{blue}O^\uparrow}$, but since there are no nontrivial events at $\bA$ and $\bB$ to act as mediators, these influences must be direct. 
Therefore, the argument goes, there must be a temporal nonlocality of causal influence, and it is this temporal nonlocality that explains the violation of screening-off. 
But from a dynamics-first point of view, this argument begins with a category mistake: it is not events but \textit{systems} that interact via the dynamics and influence each other. There is no reason that $\bA$'s lack of events relative to $\mathfrak B_2$ should prevent this system from transmitting an influence from $\bS$ to $\bO$. 
Hence the absence of events at $\bA$ and $\bB$ does not suggest temporal nonlocality, and nor does the violation of screening-off.

\section{Explanation} \label{sec:explanation}

The screening-off condition formalizes one possible account of what it means for correlations arising from a mediated causal influence to be explained by the mediator. 
Therefore, its denial by DFQT raises the question of whether that theory provides a satisfactory account of causal explanation. 

I propose that a correlation is explained by a causal mediator when it would not survive a minimal counterfactual modification of the dynamics that prevents the system from mediating the influence.  
In other words, giving the explanation means using the theory to show that in a nearby possible world in which the system did not transmit the influence, the correlation does not arise. 
This amounts to an explanation because it establishes that, in the actual world, the mediator's dynamical role is responsible for the correlation. 

To develop such an explanation of the PBR correlations, let us now describe a minimal counterfactual modification to the PBR dynamics of \cref{eq:pbr_circ} that prevents the prepared systems $\bA$ and $\bB$ from transmitting influences from $\bS$ and $\bT$ to $\bO$. The modification is naturally represented in terms of the circuit: we simply break the wires representing the causal mediators to obtain
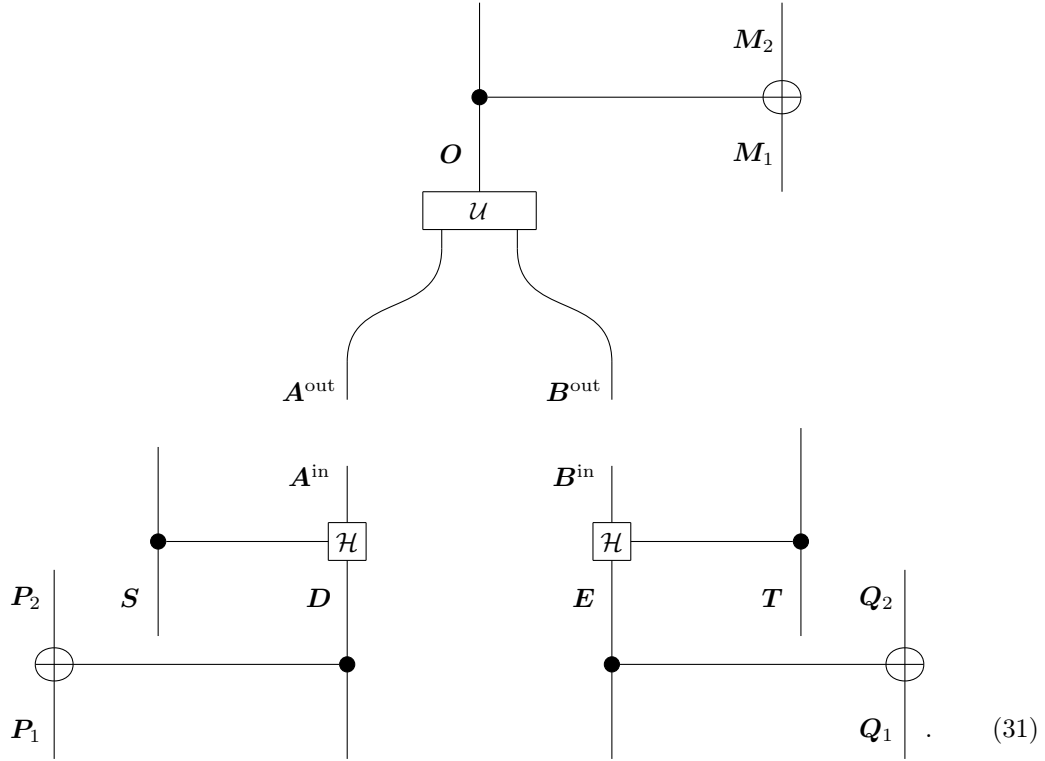
\begin{equation} \label{eq:pbr_circ_brok}
    \begin{tikzpicture}
	\begin{pgfonlayer}{nodelayer}
		\node [style=none] (0) at (9, -0.75) {};
		\node [style=none] (1) at (9, 4.5) {};
		\node [style=none] (2) at (1.25, -0.75) {};
		\node [style=none] (3) at (8.5, 4.5) {};
		\node [style=none] (4) at (9.5, 4.5) {};
		\node [style=none] (5) at (9.5, 5.5) {};
		\node [style=none] (6) at (8.5, 5.5) {};
		\node [style=none] (7) at (9, 5) {$\ch$};
		\node [style=none] (8) at (8.5, 5) {};
		\node [style=black dot] (9) at (4, 5) {};
		\node [style=none] (10) at (4, 2.5) {};
		\node [style=none] (11) at (4, 7.5) {};
		\node [style=none] (12) at (9, 5.5) {};
		\node [style=none] (13) at (1.25, 4.25) {};
		\node [style=none] (14) at (0.5, 0) {$\bP_1$};
		\node [style=none] (15) at (8.25, 3.5) {$\bD$};
		\node [style=none] (16) at (9, 9.75) {};
		\node [style=none] (17) at (11.5, 12.75) {};
		\node [style=none] (18) at (20.5, 14.25) {};
		\node [style=none] (19) at (20.5, 19.25) {};
		\node [style=none] (20) at (11, 13.25) {};
		\node [style=none] (21) at (14, 13.25) {};
		\node [style=none] (22) at (14, 14.25) {};
		\node [style=none] (23) at (11, 14.25) {};
		\node [style=none] (24) at (12.5, 14.25) {};
		\node [style=none] (25) at (12.5, 19.25) {};
		\node [style=black dot] (26) at (12.5, 16.75) {};
		\node [style=none] (27) at (12.5, 13.75) {\small $\cu$};
		\node [style=none] (28) at (11.75, 15.25) {$\bO$};
		\node [style=none] (29) at (8, 6.75) {$\bA^{\rm in}$};
		\node [style=none] (30) at (19.75, 18.25) {$\bM_2$};
		\node [style=none] (31) at (11.5, 13.25) {};
		\node [style=none] (32) at (9, 5.5) {};
		\node [style=none] (33) at (16, 9.75) {};
		\node [style=none] (34) at (13.5, 12.75) {};
		\node [style=none] (35) at (15, 6.75) {$\bB^{\rm in}$};
		\node [style=none] (36) at (13.5, 13.25) {};
		\node [style=none] (37) at (16, 5.5) {};
		\node [style=none] (38) at (21, 16.75) {};
		\node [style=none] (39) at (20, 16.75) {};
		\node [style=none] (40) at (21, 16.75) {};
		\node [style=none] (41) at (20, 16.75) {};
		\node [style=black dot] (42) at (9, 1.75) {};
		\node [style=none] (43) at (0.75, 1.75) {};
		\node [style=none] (44) at (1.75, 1.75) {};
		\node [style=none] (45) at (0.75, 1.75) {};
		\node [style=none] (46) at (1.75, 1.75) {};
		\node [style=none] (47) at (0.5, 3.5) {$\bP_2$};
		\node [style=none] (48) at (16, -0.75) {};
		\node [style=none] (49) at (16, 4.5) {};
		\node [style=none] (50) at (23.75, -0.75) {};
		\node [style=none] (51) at (16.5, 4.5) {};
		\node [style=none] (52) at (15.5, 4.5) {};
		\node [style=none] (53) at (15.5, 5.5) {};
		\node [style=none] (54) at (16.5, 5.5) {};
		\node [style=none] (55) at (16, 5) {$\ch$};
		\node [style=none] (56) at (16.5, 5) {};
		\node [style=black dot] (57) at (21, 5) {};
		\node [style=none] (58) at (21, 2.5) {};
		\node [style=none] (59) at (21, 8) {};
		\node [style=none] (60) at (16, 5.5) {};
		\node [style=none] (61) at (23.75, 4.25) {};
		\node [style=none] (62) at (20.25, 3.5) {$\bT$};
		\node [style=none] (63) at (23, 0) {$\bQ_1$};
		\node [style=none] (64) at (15.25, 3.5) {$\bE$};
		\node [style=none] (65) at (16, 5.5) {};
		\node [style=black dot] (66) at (16, 1.75) {};
		\node [style=none] (67) at (24.25, 1.75) {};
		\node [style=none] (68) at (23.25, 1.75) {};
		\node [style=none] (69) at (24.25, 1.75) {};
		\node [style=none] (70) at (23.25, 1.75) {};
		\node [style=none] (71) at (23, 3.5) {$\bQ_2$};
		\node [style=none] (72) at (9, 7) {};
		\node [style=none] (73) at (16, 7) {};
		\node [style=none] (74) at (8, 9) {$\bA^{\rm out}$};
		\node [style=none] (75) at (15, 9) {$\bB^{\rm out}$};
		\node [style=none] (76) at (16, 8.75) {};
		\node [style=none] (77) at (9, 8.75) {};
		\node [style=none] (78) at (3.25, 3.5) {$\bS$};
		\node [style=none] (79) at (19.75, 15.25) {$\bM_1$};
	\end{pgfonlayer}
	\begin{pgfonlayer}{edgelayer}
		\draw (1.center) to (0.center);
		\draw (6.center) to (3.center);
		\draw (3.center) to (4.center);
		\draw (4.center) to (5.center);
		\draw (5.center) to (6.center);
		\draw (9) to (8.center);
		\draw (11.center) to (10.center);
		\draw [in=90, out=-90, looseness=1.25] (17.center) to (16.center);
		\draw (23.center) to (20.center);
		\draw (20.center) to (21.center);
		\draw (21.center) to (22.center);
		\draw (22.center) to (23.center);
		\draw (24.center) to (25.center);
		\draw (31.center) to (17.center);
		\draw [in=90, out=-90, looseness=1.25] (34.center) to (33.center);
		\draw (36.center) to (34.center);
		\draw (26) to (38.center);
		\draw (19.center) to (18.center);
		\draw [bend left=90, looseness=1.50] (39.center) to (38.center);
		\draw [bend right=90, looseness=1.50] (41.center) to (40.center);
		\draw (42) to (43.center);
		\draw [bend right=90, looseness=1.50] (44.center) to (43.center);
		\draw [bend left=90, looseness=1.50] (46.center) to (45.center);
		\draw (2.center) to (13.center);
		\draw (49.center) to (48.center);
		\draw (54.center) to (51.center);
		\draw (51.center) to (52.center);
		\draw (52.center) to (53.center);
		\draw (53.center) to (54.center);
		\draw (57) to (56.center);
		\draw (59.center) to (58.center);
		\draw (66) to (67.center);
		\draw [bend left=90, looseness=1.50] (68.center) to (67.center);
		\draw [bend right=90, looseness=1.50] (70.center) to (69.center);
		\draw (50.center) to (61.center);
		\draw (73.center) to (65.center);
		\draw (72.center) to (32.center);
		\draw (76.center) to (33.center);
		\draw (77.center) to (16.center);
	\end{pgfonlayer}
\end{tikzpicture}.
\end{equation}
In this counterfactual circuit, there is no longer any single system called $\bA$. There are two distinct systems, $\bA^{\rm in}$ and $\bA^{\rm out}$, which respectively go into and come out of the break in the wire. The counterfactual system $\bA^{\rm in}$ ($\bA^{\rm out}$) bears the same dynamical relations to the systems below (above) the break as does $\bA$ in the actual circuit. 

Breaking the circuit in this way preserves the nontrivial algebras selected by (dual) decoherence while changing the systems to which some of them are associated. Indeed, for any circuit and $\mathfrak B$, if we break the wires corresponding to a subset of $\mathfrak B$ and define $\mathfrak B'$ as the set obtained from $\mathfrak B$ by replacing each broken $\bX$ with the corresponding $\bX^{\rm in}$ and $\bX^{\rm out}$, then we will always find that
\begin{equation}
\begin{aligned}
(\bX^{\rm in})^\uparrow_{{\rm dec}\mathfrak B'} &= \mathbb C I,
& \qquad \qquad 
(\bX^{\rm out})^\uparrow_{{\rm dec}\mathfrak B'} &= \bX^\uparrow_{{\rm dec}\mathfrak B},
\\[0.5ex]
(\bX^{\rm in})^\downarrow_{{\rm dec}\mathfrak B'} &= \bX^\downarrow_{{\rm dec}\mathfrak B},
&
(\bX^{\rm out})^\downarrow_{{\rm dec}\mathfrak B'} &= \mathbb C I,
\end{aligned}
\end{equation}
while the (dual) decohered subalgebras of the other systems of interest are left unchanged.\footnote{This is immediately obvious in the original formulation of DFQT from \citep{ormrod2024quantum}, proven by \citep{ormrod2026decoherence} to be equivalent to the present formulation. It is also not too hard to see in the present formulation.} Thus, in the present example, the decohered and dual decohered algebras for $\mathfrak B_2'$ are
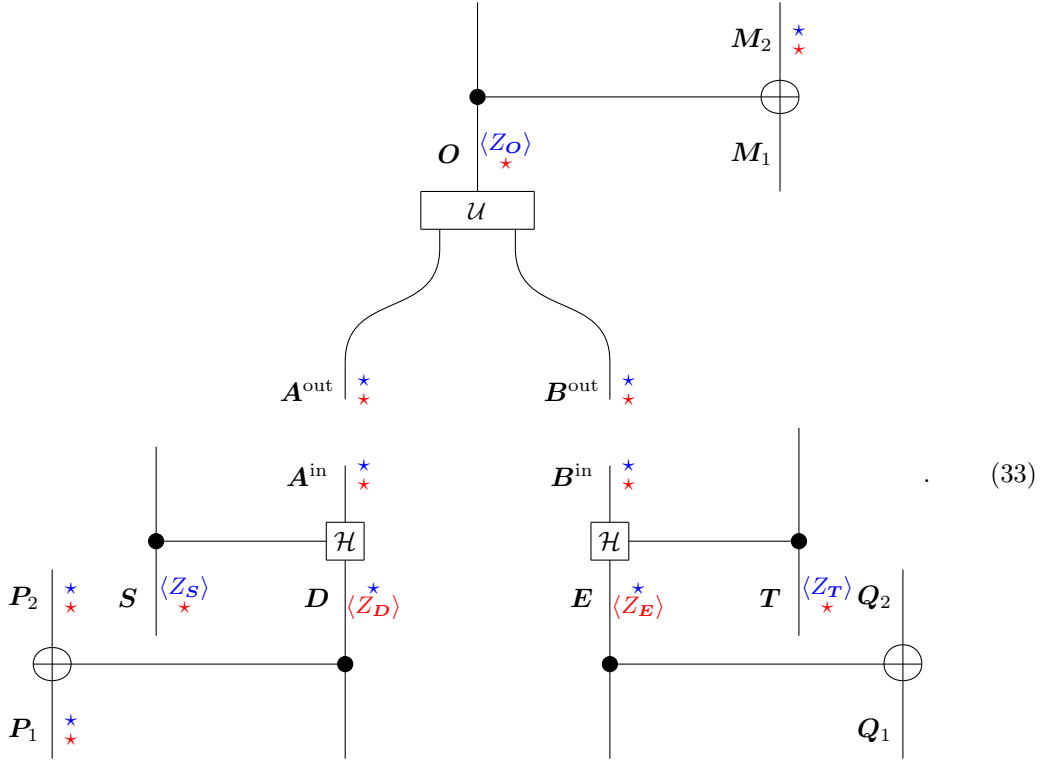
\begin{equation} \label{eq:pbr_circ_brok_dec}
    \begin{tikzpicture}
	\begin{pgfonlayer}{nodelayer}
		\node [style=none] (0) at (9, -7.5) {};
		\node [style=none] (1) at (9, -2.25) {};
		\node [style=none] (2) at (1.25, -7.5) {};
		\node [style=none] (3) at (8.5, -2.25) {};
		\node [style=none] (4) at (9.5, -2.25) {};
		\node [style=none] (5) at (9.5, -1.25) {};
		\node [style=none] (6) at (8.5, -1.25) {};
		\node [style=none] (7) at (9, -1.75) {$\ch$};
		\node [style=none] (8) at (8.5, -1.75) {};
		\node [style=black dot] (9) at (4, -1.75) {};
		\node [style=none] (10) at (4, -4.25) {};
		\node [style=none] (11) at (4, 0.75) {};
		\node [style=none] (12) at (9, -1.25) {};
		\node [style=none] (13) at (1.25, -2.5) {};
		\node [style=none] (14) at (0.5, -6.75) {$\bP_1$};
		\node [style=none] (15) at (8.25, -3.25) {$\bD$};
		\node [style=none] (16) at (9, 3) {};
		\node [style=none] (17) at (11.5, 6) {};
		\node [style=none] (18) at (20.5, 7.5) {};
		\node [style=none] (19) at (20.5, 12.5) {};
		\node [style=none] (20) at (11, 6.5) {};
		\node [style=none] (21) at (14, 6.5) {};
		\node [style=none] (22) at (14, 7.5) {};
		\node [style=none] (23) at (11, 7.5) {};
		\node [style=none] (24) at (12.5, 7.5) {};
		\node [style=none] (25) at (12.5, 12.5) {};
		\node [style=black dot] (26) at (12.5, 10) {};
		\node [style=none] (27) at (12.5, 7) {\small $\cu$};
		\node [style=none] (28) at (11.75, 8.5) {$\bO$};
		\node [style=none] (29) at (8, 0) {$\bA^{\rm in}$};
		\node [style=none] (30) at (19.75, 11.5) {$\bM_2$};
		\node [style=none] (31) at (11.5, 6.5) {};
		\node [style=none] (32) at (9, -1.25) {};
		\node [style=none] (33) at (16, 3) {};
		\node [style=none] (34) at (13.5, 6) {};
		\node [style=none] (35) at (15, 0) {$\bB^{\rm in}$};
		\node [style=none] (36) at (13.5, 6.5) {};
		\node [style=none] (37) at (16, -1.25) {};
		\node [style=none] (38) at (21, 10) {};
		\node [style=none] (39) at (20, 10) {};
		\node [style=none] (40) at (21, 10) {};
		\node [style=none] (41) at (20, 10) {};
		\node [style=black dot] (42) at (9, -5) {};
		\node [style=none] (43) at (0.75, -5) {};
		\node [style=none] (44) at (1.75, -5) {};
		\node [style=none] (45) at (0.75, -5) {};
		\node [style=none] (46) at (1.75, -5) {};
		\node [style=none] (47) at (0.5, -3.25) {$\bP_2$};
		\node [style=none] (48) at (16, -7.5) {};
		\node [style=none] (49) at (16, -2.25) {};
		\node [style=none] (50) at (23.75, -7.5) {};
		\node [style=none] (51) at (16.5, -2.25) {};
		\node [style=none] (52) at (15.5, -2.25) {};
		\node [style=none] (53) at (15.5, -1.25) {};
		\node [style=none] (54) at (16.5, -1.25) {};
		\node [style=none] (55) at (16, -1.75) {$\ch$};
		\node [style=none] (56) at (16.5, -1.75) {};
		\node [style=black dot] (57) at (21, -1.75) {};
		\node [style=none] (58) at (21, -4.25) {};
		\node [style=none] (59) at (21, 1.25) {};
		\node [style=none] (60) at (16, -1.25) {};
		\node [style=none] (61) at (23.75, -2.5) {};
		\node [style=none] (62) at (20.25, -3.25) {$\bT$};
		\node [style=none] (63) at (23, -6.75) {$\bQ_1$};
		\node [style=none] (64) at (15.25, -3.25) {$\bE$};
		\node [style=none] (65) at (16, -1.25) {};
		\node [style=black dot] (66) at (16, -5) {};
		\node [style=none] (67) at (23.25, -5) {};
		\node [style=none] (68) at (23, -3.25) {$\bQ_2$};
		\node [style=none] (69) at (9, 0.25) {};
		\node [style=none] (70) at (16, 0.25) {};
		\node [style=none] (71) at (8, 2.25) {$\bA^{\rm out}$};
		\node [style=none] (72) at (15, 2.25) {$\bB^{\rm out}$};
		\node [style=none] (73) at (16, 2) {};
		\node [style=none] (74) at (9, 2) {};
		\node [style=none] (75) at (3.25, -3.25) {$\bS$};
		\node [style=none] (76) at (9.75, -3.5) {\color{red}\small $\gen{Z_\bD}$};
		\node [style=none] (77) at (9.75, -3) {\color{blue}\small $\star$};
		\node [style=none] (78) at (1.75, -3) {\color{blue} \small $\star$};
		\node [style=none] (79) at (1.75, -6.5) {\color{blue}\small $\star$};
		\node [style=none] (80) at (4.75, -3) {\color{blue}\small $\gen{Z_\bS}$};
		\node [style=none] (81) at (4.75, -3.5) {\color{red}\small $\star$};
		\node [style=none] (82) at (9.5, 2.5) {\color{blue}\small $\star$};
		\node [style=none] (83) at (16.5, 0.25) {\color{blue}\small $\star$};
		\node [style=none] (84) at (13.25, 8.75) {\color{blue}\small $\gen{Z_\bO}$};
		\node [style=none] (85) at (21, 11.75) {\color{blue}\small $\star$};
		\node [style=none] (86) at (16.75, -3.5) {\color{red} \small $\gen{Z_\bE}$};
		\node [style=none] (87) at (16.75, -3) {\color{blue}\small $\star$};
		\node [style=none] (88) at (21.75, -3) {\color{blue}\small $\gen{Z_\bT}$};
		\node [style=none] (89) at (21.75, -3.5) {\color{red}\small $\star$};
		\node [style=none] (90) at (1.75, -7) {\color{red}\small $\star$};
		\node [style=none] (91) at (13.25, 8.25) {\color{red}\small $\star$};
		\node [style=none] (92) at (21, 11.25) {\color{red}\small $\star$};
		\node [style=none] (93) at (16.5, -0.25) {\color{red}\small $\star$};
		\node [style=none] (94) at (9.5, 2) {\color{red}\small $\star$};
		\node [style=none] (95) at (1.75, -3.5) {\color{red}\small $\star$};
		\node [style=none] (96) at (16.5, 2.5) {\color{blue}\small $\star$};
		\node [style=none] (97) at (16.5, 2) {\color{red}\small $\star$};
		\node [style=none] (98) at (9.5, 0.25) {\color{blue}\small $\star$};
		\node [style=none] (99) at (9.5, -0.25) {\color{red}\small $\star$};
		\node [style=none] (100) at (24.25, -5) {};
		\node [style=none] (101) at (19.75, 8.5) {$\bM_1$};
	\end{pgfonlayer}
	\begin{pgfonlayer}{edgelayer}
		\draw (1.center) to (0.center);
		\draw (6.center) to (3.center);
		\draw (3.center) to (4.center);
		\draw (4.center) to (5.center);
		\draw (5.center) to (6.center);
		\draw (9) to (8.center);
		\draw (11.center) to (10.center);
		\draw [in=90, out=-90, looseness=1.25] (17.center) to (16.center);
		\draw (23.center) to (20.center);
		\draw (20.center) to (21.center);
		\draw (21.center) to (22.center);
		\draw (22.center) to (23.center);
		\draw (24.center) to (25.center);
		\draw (31.center) to (17.center);
		\draw [in=90, out=-90, looseness=1.25] (34.center) to (33.center);
		\draw (36.center) to (34.center);
		\draw (26) to (38.center);
		\draw (19.center) to (18.center);
		\draw [bend left=90, looseness=1.50] (39.center) to (38.center);
		\draw [bend right=90, looseness=1.50] (41.center) to (40.center);
		\draw (42) to (43.center);
		\draw [bend right=90, looseness=1.50] (44.center) to (43.center);
		\draw [bend left=90, looseness=1.50] (46.center) to (45.center);
		\draw (2.center) to (13.center);
		\draw (49.center) to (48.center);
		\draw (54.center) to (51.center);
		\draw (51.center) to (52.center);
		\draw (52.center) to (53.center);
		\draw (53.center) to (54.center);
		\draw (57) to (56.center);
		\draw (59.center) to (58.center);
		\draw (50.center) to (61.center);
		\draw (70.center) to (65.center);
		\draw (69.center) to (32.center);
		\draw (73.center) to (33.center);
		\draw (74.center) to (16.center);
		\draw (66) to (100.center);
		\draw [bend left=90, looseness=1.75] (67.center) to (100.center);
		\draw [bend right=90, looseness=1.50] (67.center) to (100.center);
	\end{pgfonlayer}
\end{tikzpicture}.
\end{equation}

While breaking the circuit leaves the set of possible emergent events essentially unchanged, it does alter their probabilities. Indeed, our circuit now comes in two disconnected fragments: the fragment below the breaks and the one above. It follows that
\begin{equation} \label{eq:set_out_no_cor}
      {\rm Prob}_{\mathfrak B_2'} ( {\color{blue}O^\uparrow}| {\color{red} D^\downarrow} = {\color{red} E^\downarrow}  = {\color{red}0 },{\color{blue} S^\uparrow, T^\uparrow})  
      =  
      {\rm Prob}_{\mathfrak B_2'} ( {\color{blue}  O^\uparrow}|  {\color{red} D^\downarrow} = {\color{red} E^\downarrow} = {\color{red}0 } ), 
\end{equation}
in contrast with the correlated conditional probability distribution from \cref{eq:pbr_conditional}.

This shows that, in a nearby possible world where $\bA$ and $\bB$ do not transmit influences from $\bS$ and $\bT$ to $\bO$, there is no correlation between the setting and outcome events. It follows that the correlation in the actual world is explained by the mediating activity of $\bA$ and $\bB$. 

The strategy here for explanation by causal mediation generalizes beyond the PBR model. Indeed, let $\bX$, $\bY$, and $\bZ$ be arbitrary systems in an arbitrary unitary circuit. In the quantum causal modelling framework of \citep{Allen_2017, barrett2020quantum}, one system is said to influence another through a unitary transformation precisely if their pulled-back operator algebras do not commute.\footnote{As shown in \citep{ormrod2023causal}, this definition of influence is equivalent to the possibility of an agent with access to the cause system signalling to an agent with access to the effect system.}
Hence the statement that the influence from $\bX$ to $\bZ$ is fully mediated by $\bY$ is naturally taken to mean that in the counterfactual circuit in which $\bY$ is broken, the pulled-back algebras $\tilde \bX^{\rm count}$ and $\tilde \bZ^{\rm count}$ commute.\footnote{Indeed, this is precisely the circumstance under which the quantum causal modelling framework attributes the unitary circuit a causal structure represented by the graph $\bX \rightarrow \bY \rightarrow \bZ$.} If they do, then there is a factorization $\ch^{\rm count} = \ch_\bX \otimes \ch_{\bX}'$ of the total Hilbert space of the counterfactual circuit such that $\tilde \bX^{\rm count} =\cl(\ch_\bX) \otimes I_{\bX'}$ and $\tilde \bZ^{\rm count} \subseteq I_\bX \otimes \cl(\ch_{\bX'})$. It then follows from DFQT's probability rule that, in a possible world in which the causal connection at $\bY$ is broken, events at $\bX$ and $\bZ$ are uncorrelated. Therefore, the correlation between $\bX$- and $\bZ$-events in the actual world is explained by $\bY$'s mediating role. 

The argument generalizes straightforwardly to sets of systems $\{\bX_i\}_{i=1}^{N_\bX}$ and $\{\bZ_i\}_{i=1}^{N_\bZ}$ that are causally connected only due to influences from the former to the latter that are fully mediated by $\{\bY_i\}_{i=1}^{N_\bY}$, assuming that there are no influences between the various $\bX_i$ or between the various $\bZ_i$.
When there are such intra-set influences, the same argument fails because the combined algebras $\bigvee_i \tilde \bX_i^{\rm count}$ and $\bigvee_i \tilde \bZ_i^{\rm count}$ may be nonfactors, i.e.\ algebras that are not isomorphic to $\cl(\ch)$ for any Hilbert space $\ch$. 
However, the quantum causal modelling framework provides a natural strategy for generalizing the argument further: applying Theorem 4.10 of \citep{barrett2020quantum} to DFQT. This theorem shows under general conditions that if the influence from $\{\bX_i\}_{i=1}^{N_\bX}$ to $\{\bZ_i\}_{i=1}^{N_\bZ}$ is fully mediated by $\{\bY_i\}_{i=1}^{N_\bY}$, then correlations between classical outcomes of interventions at $\{\bX_i\}_{i=1}^{N_\bX}$ and $\{\bZ_i\}_{i=1}^{N_\bZ}$ do not survive when one intervenes on each $\bY_i$ to break the causal connection. Using quantum causal models to develop a general theory of causal explanation in DFQT lies beyond our current scope, but provides a worthwhile direction for future work.

We conclude that a shift to a dynamics-first ontology does not mean abandoning causal explanation, but switching to an explanatory strategy based on counterfactual modifications to the dynamics rather than on screening off by physical states.
Let us end our discussion of causal explanation by reflecting on an analogy between the present response to PBR and a possible response to Bell's theorem: maintain that the Bell correlations should be explained by a common cause while denying that the common cause screens them off. In both cases, the point is not to deny the need for causal explanation, but to deny that screening-off conditions (and more generally, the Markov condition of classical causal modelling \citep{spirtes2000causation, pearl2009causality}) are appropriate formalizations of causal explanation in the quantum setting. This analogy poses a challenge for those who are sympathetic to this response to Bell and yet regard PBR as a knock-down argument for $\psi$-ontology: if screening-off can fail for common causes, why not causal mediators?

\section{Is this $\psi$-epistemology?} \label{sec:epistemic}

According to DFQT, the pure quantum state assigned to a system does not, in general, supervene on physical facts about the system. 
Its role is to encode knowledge about the likelihood of possible events given other events that are known to have happened. 

In this sense, the role of the quantum state in DFQT is epistemic. 
However, this view of the quantum state diverges from the specific version of $\psi$-epistemology formulated in \citep{harrigan2010einstein}.
There, the guiding analogy was that the quantum state is like a probability distribution, describing degrees of belief about the physical properties of the system to which the probability distribution is assigned. 
But in our model of PBR, the quantum state does not describe degrees of belief about the physical properties of the system to which it is assigned; indeed, that system has no nontrivial properties relative to the set of systems of interest being considered. Hence DFQT arguably does not capture the spirit of Harrigan-Spekkens $\psi$-epistemology.

However, from a broader $\psi$-epistemic point of view, this may be an advantage. Indeed, a common and rather unfortunate misconception about the $\psi$-epistemic research programme associated with \citep{harrigan2010einstein, Spekkens_2007} is that it takes the particular definition of $\psi$-epistemology from \citep{harrigan2010einstein} to be satisfactory. To the contrary, one of the central goals of the programme is to develop an alternative, nonclassical notion of $\psi$-epistemology capable of reconciling quantum phenomena with the formulation of Leibniz's principle defended in \citep{spekkens2019ontological}. 

This work provides a proposal for what the required shift in the concept of $\psi$-epistemology might be, grounded in the shift to a dynamics-first ontology and exemplified by DFQT. If a system's dynamical role does not reduce to facts about its state space, then knowledge associated with a system does not reduce to knowledge about its state. The quantum state of a system encodes knowledge about the likelihood of possible events grounded in both other events, which may or may not be associated with the system to which the state is assigned, and in the primitive dynamical capability of the system.

\section{Conclusion} \label{sec:conc}

A correlation requires an explanation, and, if dynamics reduce to counterfactual dependence relations between states, then it is natural to explain it by showing that a state screens it off. 
But if a system's ability to transmit influences is not tied to its possession of a state, then one can still explain a correlation by showing that it would not survive the counterfactual destruction of the relevant causal connection. As DFQT demonstrates, the PBR correlations can be so explained while maintaining a broadly epistemic reading of quantum states.

Proponents of state-first ontology may continue to view the PBR theorem as a powerful argument for the reality of the quantum state. 
But suppose that one's prior commitment is not to a state-first ontology, but to an ontology without the quantum state. 
PBR is then naturally seen as an argument against state-first ontology, suggesting a shift to the dynamics-first approach. 
Thus, for critics of $\psi$-ontology, PBR points not to the reality of the quantum state, but to a vision of reality in which dynamics are more fundamental than states.

\section*{Acknowledgements} This work was inspired by an open discussion at the Perimeter Institute's 100 Years of Quantum conference and stimulated by an invitation to Chapman University's Status of the Quantum State workshop; it is a pleasure to thank the organizers of both events. In addition, this work has benefited from helpful comments from Wayne Myrvold on a draft of the manuscript, and from useful discussions with Jonathan Barrett, Matt Leifer, Tein van der Lugt, Wayne Myrvold, Robert Spekkens, and Y{\`\i}l\`e Y\={\i}ng. Finally, I gratefully acknowledge the support of the Bruce H. Mitchell Foundation, whose funding helped make this work possible.

\bibliographystyle{plainnat}
\bibliography{refs}

\end{document}